\documentclass[11pt, a4paper, logo, copyright]{customclass}

\title{A Quantum Circuit Framework for Protein Ensemble-Level Energetics}

\correspondingauthor{Pratik Patil (p.patil@uos.ac.uk), Bhushan Bonde (b.bonde@uos.ac.uk, bhushan.bonde@evotec.com) both contributed equally for the work.}

\begin{abstract}
    Proteins occupy heterogeneous free-energy landscapes, where broad, high-entropy ensembles converge toward compact, low-energy basins containing multiple sub-states. Molecular dynamics can capture these landscapes at atomic resolution, but exhaustively sampling rare transitions remains computationally demanding. Furthermore, most quantum approaches to protein modelling target only a single optimal structure, leaving the energetic heterogeneity of the full ensemble largely unexplored. We introduce a residue-level, gate-based quantum circuit framework for coarse-graining protein thermodynamics. Each amino acid is represented as a two-state qubit with stabilised and excited solvation states, initialised from solvation energetics. Following this, a structure-informed entanglement block of parametrised controlled gates encodes covalent bonds and non-covalent contacts across the residue network. The circuit is then measured $\approx 10^6$ times, yielding an ensemble of binary thermodynamic microstates, from which we compute energy distributions, residue-level statistical couplings, energetic sensitivities, and information gain relative to total free energy. We showcase the comparative results for Trp-cage miniprotein (PDB:1L2Y) and the disulfide-stabilised chimera (PDB:9GDL), which comprises a structured, folding-funnel-like energy distribution for 1L2Y. Moreover, comparison with 9GDL reveals clear shifts in global energy and residue-level profiles, localising the residues driving stabilisation and ensemble reorganisation. Additionally, dimer- and trimer-level couplings resolve both direct and interaction-mediated dependencies, showing the circuit captures network-level, not just isolated, correlations. The framework moves quantum protein modelling beyond single-structure optimisation toward interpretable, ensemble-level characterisation.
\end{abstract}


\renewcommand{\today}{}

\usepackage[utf8]{inputenc} 
\usepackage{subcaption} 
\usepackage{changepage} 
\usepackage{wrapfig}
\usepackage{longtable}
\usepackage{braket}
\usepackage{cite}
\usepackage{IEEEtrantools}
\usepackage{chngcntr}
\usepackage{verbatim}
\usepackage{graphicx}
\usepackage[nameinlink]{cleveref}
\usepackage{bbm}
\usepackage{floatrow}
\usepackage{booktabs}   
\usepackage{tabularx}
\usepackage{makecell} 
\usepackage{array}
\usepackage{cancel}
\usepackage{dblfloatfix} 

\theoremstyle{definition}

\providecommand{\cellbox}[1]{\parbox[t]{\linewidth}{\raggedright#1}}

\usepackage{xcolor} 

\definecolor{darkgreen}{RGB}{0, 102, 0}

\author[1,2]{Pratik Patil$^\dagger$}
\author[1,2]{Bhushan Bonde$^\dagger$ }
\author[1,3]{Bhaskar Choubey}
\affil[1]{ Digital Futures Institute, University of Suffolk, Ipswich, UK }
\affil[2]{ In-silico R\&D, Evotec, Abingdon, Oxfordshire, UK }
\affil[3]{ University of Siegen, Siegen, Germany}
\date{July 2026}

\begin{document}
\maketitle
\bstctlcite{BSTcontrol}
\newpage

\setcounter{figure}{0}
\counterwithin{figure}{section}

\setcounter{table}{0}
\counterwithin{table}{section}

\setcounter{equation}{0}
\counterwithin{equation}{subsection}

\section{Introduction}

Proteins function as dynamic ensembles of transitioning conformational states \cite{HenzlerWildman2007,Boehr2009,Peters2024ProteinLandscape}, their behaviour is commonly described by a rugged free-energy landscape in which broad, high-entropy ensembles converge towards compact low-energy basins containing multiple metastable sub-states \cite{Onuchic1997,Wolynes1995,Dill2012ProteinFolding}. Resolving these ensembles is important for understanding folding, molecular recognition, protein--protein interactions, and allosteric regulation \cite{Munoz07, Boehr2009, Motlagh2014}. However, experimental observables are often sparse or ensemble-averaged, classical methods such as molecular dynamics simulations can remain limited by the computational cost of crossing high free-energy barriers and sampling rare transitions \cite{Bonomi2017EnsembleDetermination,Karplus2002MolecularDynamics,Valsson2016RareEvents}. Alternatively, quantum computing approaches have been proposed as potentially promising tools for investigating such problems \cite{Bonde2024, Ghamari2022QuantumTransitions, Ghamari2024RareProteinTransition, Babbush2014LatticeHeteroPolymer, Robert2021, li2025quantumalgorithmproteinstructure}. However, these methods remain in a nascent stage and require sustained development and further methodological advances. Moreover, Quantum Computing approaches to protein modelling have primarily treated folding as an optimisation problem, targeting a lowest-energy structure or a set of low-energy lattice conformations \cite{PerdomoOrtiz2012QA,boulebnane2023peptideQAOA,li2025quantumalgorithmproteinstructure}. Hybrid quantum-annealing methods have additionally been explored to sample transition rare pathways between predefined molecular states \cite{Ghamari2022QuantumTransitions,Ghamari2024RareProteinTransition}. Furthermore, Quantum approximate optimisation algorithm (QAOA) has been found to be ineffective for sampling low-energy conformations \cite{boulebnane2023peptideQAOA}. Closer to the present aim, a preprint manuscript from 2025 reports a turn-based encoding on a face-centred cubic lattice with eighteen degrees of freedom has been used to predict peptide backbones and to construct a free-energy surface from the sampled bitstrings, solved with a variational quantum eigensolver using conditional value-at-risk (CVaR) \cite{kannan2025QFreeEnergy}. Therein, each amino acid is coarse-grained to a single bead centred on its C$\alpha$ atom, and inter-residue interactions are represented by the Miyazawa--Jernigan knowledge-based contact potential, however, we note that the hamiltonian has only pairwise contact energies but no single-residue term ($\sum_i(.)_i$ form). This implies that a residue's burial or exposure preference is expressed only through the contacts it realises and no intrinsic transfer free energy is assigned to the residue itself; side-chain orientation and explicit solvation are likewise absent, and are identified by the authors as future work. Moreover, the reported surface is a population map over contact energy and radius of gyration constructed from the measured bitstrings. Because those bitstrings are drawn from a variationally optimised state under a risk-measure objective that by construction favours the low-energy tail, the sampling weight is not established as a Boltzmann distribution over the encoded objective. The resulting ensemble concentrates on the basins reachable from a single converged parameter setting, so that metastable or functionally relevant intermediates lying in competing basins may remain under-sampled \cite{kannan2025QFreeEnergy}. Further, while writing this manuscript, a non-iterative quantum-sampling architecture was reported that generates candidate peptide conformations through a single forward Hamiltonian evolution and reconstructs an approximate energy landscape from the resulting structural ensemble \cite{zhang2026quantumsamplingarchitectureprotein}. Therein, each peptide is represented as a self-avoiding walk on a tetrahedral lattice, with pairs of qubits encoding backbone-turn directions and auxiliary qubits declaring residue contacts. Two consequences of this representation are relevant here. First, contacts are admissible only at odd sequence separations of five or more residues, a restriction the authors attribute to the bipartite geometry of the lattice; residue pairs at even separation therefore have no interaction term in the Hamiltonian irrespective of their proximity in the folded structure. Second, burial preference is applied as a bias on the contact register rather than on solvent exposure, and that register is auxiliary: multiple contact-qubit patterns correspond to the same backbone fold, and a separate burial criterion is imposed classically when candidate structures are ranked. Side-chain coordinates are likewise absent from the quantum representation, entering through fixed residue-dependent coefficients whose values are not reported, with all-atom side chains rebuilt after sampling. The authors note that the resulting landscapes are two-dimensional projections and should be interpreted as approximate structural energy maps rather than thermodynamic free-energy surfaces \cite{zhang2026quantumsamplingarchitectureprotein}.

Herein, we introduce a residue-level thermodynamic framework for gate-based quantum sampling. Each amino acid is represented as a two-state qubit whose initial excitation propensity is determined by residue solvation energetics \cite{FaucherePliska1983,Tien2013}. Further, the Parametrised interaction block in our quantum circuit encode covalent and non-covalent interactions, and applies them as parametrised controlled rotations. This introduces correlations across the residue-interaction network, producing a joint distribution over stabilised and excited residue states. The circuit is measured repeatedly for $N$-times, to generate binary configurations (bit-string for amino-acid sequence) of thermodynamic microstates. From this binary configurations we apply naive hamiltonian operator to derive energy distributions, residue stability profiles, energetic sensitivities, and direct or interaction-mediated statistical couplings. We showcase the framework by comparing the results for Trp-cage structures: 1L2Y and 9GDL.

\section{Methods}
\label{sec:methods}

\begin{figure}[ht]
    \centering
    \includegraphics[width=\linewidth]{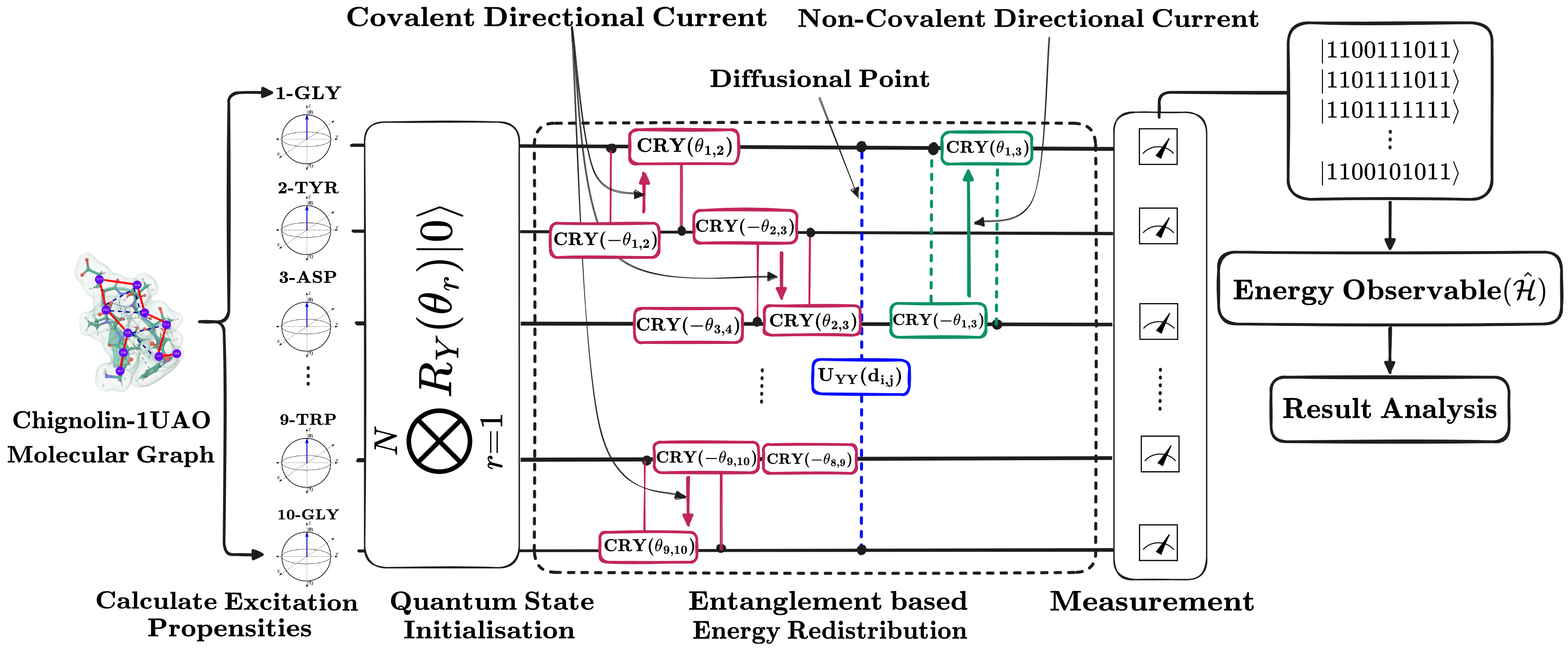}
    \caption{\textbf{Quantum residue-ensemble workflow.}
    Each residue is mapped to a qubit.
    Qubits are initialised from solvent- and hydrophobicity-modulated excitation propensities. Pairwise interaction updates redistribute excitation on covalent and non-covalent edges. Repeated measurements produce a microstate ensemble for energy and information analysis.}
    \label{fig:quantum-circuit}
\end{figure}

We construct residue-level ensemble statistics from a static structure of protein molecule available from protein data bank (PDB) format. The workflow comprises six stages:
\begin{enumerate}
\item Input PDB structure: map each amino-acid (AA) in the protein to a qubit on the quantum circuit 
\item Calculate AA solvent exposure and local excitation energetics in the protein structure
\item Prepare an initial quantum state from the AA excitation propensity
\item Construct AA interaction graph
\item Calculate the qubit entanglement parameters from the interaction graph based on coupling strength and energetic potential
\item Measure the quantum circuit to generate the ensemble statistics
\item Ensemble analysis: Downstream clustering (AA dimer/trimer coupling analysis)
\end{enumerate}

Please refer to Fig.~\ref{fig:quantum-circuit} as a visual representation of the workflow, to be read from left to right, the following subsections describe each stage in detail.

\subsection{Residue representation and solvent exposure}
\label{sec:sasa}

Given a static PDB structure of a protein molecule with $N$ amino acid residues (AAs), the residues are mapped one-to-one with the qubits on the quantum circuit $r\in\{1,\dots,N\}$. For each residue, the solvation state of a residue was calculated considering Solvent Accessible Surface Area (SASA). The SASA for each residue is calculated using FreeSASA tool \cite{Mitternacht2016FreeSASA} and then normalised by residue-specific theoretical maximum accessible surface area (MAXASA) values \cite{Tien2013}:
\begin{align}
    \alpha_r
    =
    \min\!\left(0.999,\;\frac{\mathrm{SASA}_r}{\mathrm{MAXASA}_r}\right),
    \label{eq:alpha}
\end{align}
where $\mathrm{SASA}_r$ is the solvent-accessible area of residue $r$ and $\mathrm{MAXASA}_r$ is the corresponding reference maximum. The parameter $\alpha_r\in[0,0.999]$ therefore quantifies the relative solvent exposure of each residue, with values near zero indicating buried positions and values near unity indicate full residue exposure.

\subsection{Local energetics and excitation propensity}
\label{sec:local-energy}

Further, each residue were assigned two quantum states characterised as: stabilised (ground) state with energy $E_{0,r}$ and an excited state with energy $E_{1,r}$. The excited state and ground state were represented w.r.t. to their free transfer energies from water to bulk octanol \cite{FaucherePliska1983}. The rationale being, hydrophic AAs are stable in a non-polar solvent such as bulk octanol characterising the buried nature of hydrophobic core in the proteins. While the polar residues are stable in polar solvent characterising the solvent exposure of hydrophilic residues. Their relative transfer free energies, hence can be used to represent the excited state and ground states of these residue. For example, a residue like Tryptophan with water exposure is considered in excited state, while a residue like Lysine is considered stable with water exposure. This analogy establishes the solvation characteristics of each residue and forms the basis for the two-level quantum system considered in the method. All the free transfer energies are expressed relative to the transfer free energy of glycine from water to bulk octanol \cite{FaucherePliska1983}, table~\ref{tab:transfer_energy_refs} provides corresponding values for reference.

The intrinsic excitation gap ($\Delta G_r$) for residue $r$ is given by:

\begin{align}
    \Delta G_r = E_{1,r} - E_{0,r}.
    \label{eq:dG}
\end{align}
where $E_{1,r}$ is the excited state energy and $E_{0,r}$ is the ground state energy. Further, the effective excitation gap ($\Delta G_r^{\mathrm{eff}}$) defined in eq.~\ref{eq:dG_eff_piecewise} describes the solvation state of the residue in the snapshot structure observed. 

\begin{align}
    \Delta G_r^{\mathrm{eff}}
    =
    \begin{cases}
    \Delta G_r\,(1-\alpha_r), & \mathrm{KD}_r > 0.5, \\[4pt]
    \Delta G_r\,(1-\max(\alpha_r,0.1)), & -0.5 < \mathrm{KD}_r \le 0.5, \\[4pt]
    \Delta G_r\,\alpha_r, & \mathrm{KD}_r \le -0.5.
\end{cases}
\label{eq:dG_eff_piecewise}
\end{align}

where $\mathrm{KD}_r$ denote the Kyte--Doolittle hydrophobicity index of residue $r$ \cite{KyteDoolittle1982} and $\alpha_r$ is the relative SASA from eq.~\ref{eq:alpha}. $\mathrm{KD}_r > 0.5$) represent the Hydrophobic residues case, while $\mathrm{KD}_r \le -0.5$ represent Hydrophilic residues solvation characteristics. The intermediate branch enforces a minimum exposure floor of $\alpha_r \ge 0.1$ to prevent degeneracy for neutral residues such as Ser/Gly with near-zero computed SASA.

The excitation propensity $p_r$ is then obtained from a two-level Boltzmann partition:

\begin{align}
    p_r
    &=
    \frac{\exp\!\left(-\beta_T \,\Delta G_r^{\mathrm{eff}}\right)}
         {1+\exp\!\left(-\beta_T \,\Delta G_r^{\mathrm{eff}}\right)},
    \qquad
    \beta_T = \frac{1}{R T},
    \label{eq:pr_boltz}
\end{align}

where $T$ is the absolute temperature and $R$ is the gas constant. Note that the free transfer energies reported by Fauchère and Pliska were measured at room temperature ($\approx 300k$)  accordingly, the temperature parameter used here is set to $\approx 300k$. That said, this should be viewed as a parametrisation rather than a hard constraint \cite{FaucherePliska1983}.

\subsection{Quantum encoding of propensities}
\label{sec:encoding}

The residue excitation propensities from eq.~\ref{eq:pr_boltz} were then encoded onto quantum circuit using single-qubit ($R_Y$) rotations on $\ket{\psi_0}$ to form the initial quantum state ($\ket{\psi_0}$) of the system, given by:

\begin{align}
    \ket{\psi_0}
    =
    \bigotimes_{r=1}^{N} R_Y(\theta_r)\ket{0},
    \qquad
    \theta_r = 2\arcsin\!\sqrt{p_r}.
    \label{eq:ry_init}
\end{align}

Equivalently, the quantum state of qubit $r$ represented as $\ket{\psi_{r}}$) can be expressed as superposition of $\ket{0}$ and $\ket{1}$:
\begin{align}
    \ket{\psi_{r}}
    &=
    \sqrt{1-p_r}\ket{0}
    +
    \sqrt{p_r}\ket{1}.
\end{align}
At this stage, measurement of qubit $r$ would produce ($\ket{1}$) with $p_r$ probability and $\ket{0}$ with $1-p_r$ probability. Here, the initialisation contains only single-qubit operations, and hence the initial state is a separable product state and contains no inter-residue correlations, see appex.~\ref{app:encoding_angle} for more detail on encoding. The following subsections~\ref{subsec:kij} and~\ref{subsec:entanglement} establish the residue interactions based modulation to quantum state of the system.

\subsection{Structure-derived residue coupling}
\label{subsec:kij}

The protein was represented as a residue graph ($\mathcal{G}=(\mathcal{V},\mathcal{E})$), where each node ($\mathcal{V}$) represents one residue and each edge ($\mathcal{E}$) represents amino-acid interaction. The edge set contained peptide-backbone connections and selected non-covalent contacts. For every residue pair $(i,j)$, a structural stiffness score $S_{ij}$ was calculated from backbone connectivity ($\mathrm{BB}$), disulfide bonding ($\mathrm{SS}$), hydrogen bonding($\mathrm{HB}$), salt-bridge ($\mathrm{salt}$) formation, and van der Waals ($\mathrm{vdW}$) packing:

\begin{align}
    S_{ij}
    =
    w_{\mathrm{BB}}\kappa^{\mathrm{BB}}_{ij}
    +
    w_{\mathrm{SS}}\kappa^{\mathrm{SS}}_{ij}
    +
    w_{\mathrm{HB}}\kappa^{\mathrm{HB}}_{ij}
    +
    w_{\mathrm{salt}}\kappa^{\mathrm{salt}}_{ij}
    +
    w_{\mathrm{vdW}}\kappa^{\mathrm{vdW}}_{ij}.
    \label{eq:sij}
\end{align}

Here, ($\kappa^{x}_{ij}$) is the pairwise contribution associated with interaction class ($x$), and ($w_x$) is its corresponding weighting coefficient. This graph-based construction follows the general representation used in elastic- and Gaussian-network descriptions, in which structural contacts define effective couplings between residue degrees of freedom \cite{Tirion1996,Haliloglu1997,Atilgan2001,Bahar2006}. Moreover, the above structural stiffness score ($S_{ij}$) was further modulated by frequency matching ($\Phi_{ij}$) and mass compatibility ($\Lambda^{\mathrm{mass}}_{ij}$), giving the final coupling score ($K_{ij}$):
\begin{align}
    K_{ij}
    =
    S_{ij}\Phi_{ij}\Lambda^{\mathrm{mass}}_{ij}.
    \label{eq:kij}
\end{align}

Importantly, positive couplings ($K_{ij}$) were normalised within each protein. Residue pairs with ($K_{ij}\leq0$) were not included in the interaction graph. Definitions of the contact descriptors, geometric cutoffs, weighting coefficients, frequency-matching function, and mass-compatibility function are provided in Appendix~\ref{app:kij}. In summary, the coefficient ($K_{ij}$) suggest how strongly the state of two amino-acids are coupled that can be further used for entanglement parametrisation.

\subsection{Quantum circuit entanglement block}
\label{subsec:entanglement}

After applying the initial operation on qubits described by eqn~\ref{eq:ry_init}, we apply the entanglement block ansatz to the circuit, that captures the residue-level interactions. Herein, circuit entanglement is interpreted as a computational representation of structure-mediated statistical dependence between residue excitation states. The circuit parameters for the entanglement block depends on effective thermodynamic drive between residues, we considered an effective local excitation potential ($\mu_i$) for each residue ($R_i$), defined as

\begin{align} 
    \mu_i &= \beta p_i \Delta G_i  , \label{eq:mu_eff} 
\end{align}

where \(\Delta G_i\) is the residue-level excitation or transfer free-energy scale, \(p_i\) is the probability that residue \(i\) is initially in the excited state. The effective local excitation potential ($\mu_i$) measures a residue's solvation excited state potential. Further, for each ordered residue pair \(i\rightarrow j\), the excitation transfer flux ($\nu_{i \rightarrow j}$) is then considered as 

\begin{align} 
    \nu_{i\rightarrow j} &= \frac{\mu_i-\mu_j} {|\mu_i|+|\mu_j|}, \label{eq:directed_drive} 
\end{align} 

 Here, excitation transfer happens from a residue with larger effective excitation potential to a residue with smaller effective excitation potential. The denominator is normalised to the local energy scale of the pair, so that \(\nu_{i\rightarrow j}\) represents a relative imbalance rather than an unbounded energy difference.

The physical motivation for this construction comes from vibrational energy transfer (VET) in proteins. Experimental and computational studies have shown that excess vibrational energy can propagate through both covalent backbone pathways and non-covalent contacts, including hydrogen bonds, salt bridges, and stacking contacts \cite{Deniz2021}. Such energy-flow pathways have also been linked to allosteric communication and energy dissipation in proteins \cite{Leitner2008EnergyFlow,Li2014AnisotropicEnergyFlow,Deniz2021}. Therefore, the present model treats residue excitation transfer as a coarse-grained analogue of structure-mediated energy redistribution rather than as a microscopic rate expression. In this interpretation, \(K_{ij}\) and \(\nu_{i\rightarrow j}\) have distinct roles. The coupling factor \(K_{ij}\) encodes the structural conductance of the pair: it determines whether residues \(i\) and \(j\) are connected through backbone continuity, disulfide crosslinks, hydrogen bonding, salt bridges, or short-range packing contacts. The drive \(\nu_{i\rightarrow j}\) encodes the directional energetic bias. Their product therefore gives a structure-conditioned transfer tendency, 
\begin{align}                      
    K_{ij}\nu_{i\rightarrow j}        &\equiv \text{structural coupling} \times \text{directional excitation bias}.
    \label{eqn:pre-delta-p}
\end{align} 

The residue excitation is largely governed by how tightly the water molecules are coupled to the amino acid residues, and an increase in local excitation of residues loosen up the residue-water coupling and thus change the local solvation stability \cite{Singh2026,Deniz2021}. Based on this fact, we model the directed transfer probability for the entanglement block from eqn.~\ref{eqn:pre-delta-p} as:

\begin{align} 
    \Delta P_{i\rightarrow j} &= \frac{ 2K_{ij}\nu_{i\rightarrow j} }{ \sum_{k\in\mathcal{N}(i),\mathcal{N}(j)}K_{ik} }. \label{eq:directed_transfer_probability}
\end{align} 

This transfer probability as in en~\ref{eq:directed_transfer_probability} represents the statistical dependence between the residues. It is characterised by both, a structural pathway and a positive directional drive, represented by \(K_{ij}\nu_{i\rightarrow j}\). The normalisation over \(\sum_{k\in\mathcal{N}(i)}K_{ik}\) distributes the outgoing transfer tendency of residue \(i\) across its structurally connected neighbours, preventing highly connected residues from transferring excitation without accounting for competition among local pathways. Further, the probability-like interaction parameter was mapped to a controlled ($CRY$) rotation gate, that forms the fundamental gate for the entanglement block. The following eq.~\ref{eq:cry_relation} expresses the entanglement operation on top of initial quantum state as in eqn.~\ref{eq:ry_init}.
\begin{align}
    \vartheta_{i\rightarrow j}
    &=
    2\arcsin
    \sqrt{\Delta P_{i\rightarrow j}},
    \label{eq:cry_theta}
\end{align}
\begin{align}
    U_{i\rightarrow j}
    =
    \operatorname{CRY}_{i\rightarrow j}
    \left(
    \vartheta_{i\rightarrow j}
    \right);U_{j\rightarrow i}
    =
    \operatorname{CRY}_{j\rightarrow i}
    \left(
    -\vartheta_{i\rightarrow j}
    \right)
    \label{eq:cry_relation}
\end{align}
Residue $R_i$ acts as the control qubit and residue $R_j$ as the target, that suggests the propensity flows from $i \rightarrow j$. Consequently, the state of residue $j$ is updated conditionally when residue $i$ occupies the excited state. Because the interaction gates generally do not commute (see appx.~\ref{app:gate_ordering}), their order can affect the sampled distribution. We scheduled gate operations ordering as stronger couplings before weaker ones based on $K_{ij}$. This reduces gate conflicts and limits disruption of correlations introduced by dominant interactions. Nonetheless, entanglement ordering remains a modelling choice and should remain open to further research. 

Additionally, to model a light entropy of the solvation bath, we introduce a second, symmetric coupling for residue pairs whose thermodynamic drive is weak. For each ordered pair $(i,j)$ in the scheduled interaction list, we take the effective drive proxy $\Delta^{\mathrm{eff}}_{i\rightarrow j}$ to be the directed potential imbalance $\nu_{i\rightarrow j}$ of Eq.~\ref{eq:directed_drive} where available, falling back to the structural transfer probability $\Delta P_{i\rightarrow j}$ of Eq.~\ref{eq:directed_transfer_probability} otherwise. When this drive is small,
\begin{align}
    \left|\Delta^{\mathrm{eff}}_{i\rightarrow j}\right| \;\le\; \tau_{YY},
    \label{eq:yy_threshold}
\end{align}
residues $i$ and $j$ are treated as thermodynamically near-degenerate: with no clear excitation hierarchy between them, the directed $\operatorname{CRY}$ construction of Eq.~\ref{eq:cry_relation} is not appropriate, since it would impose an arbitrary control-target asymmetry on a pair that is, physically, exchanging energy on comparable footing through shared solvent fluctuations rather than along a defined gradient. For such pairs we replace the $\operatorname{CRY}$ layer with a single, symmetric $\operatorname{IsingYY}$ gate,
\begin{align}
    U^{YY}_{ij}\!\left(\vartheta^{YY}_{ij}\right) = \exp\!\left(-i\,\frac{\vartheta^{YY}_{ij}}{2}\,\hat{Y}_i\hat{Y}_j\right), \qquad
    \vartheta^{YY}_{ij} = \operatorname{clip}\!\left(2\arcsin\!\sqrt{\left|\Delta^{\mathrm{eff}}_{i\rightarrow j}\right|},\; -\Theta_{\mathrm{cap}},\, \Theta_{\mathrm{cap}}\right),
    \label{eq:isingyy}
\end{align}
using the same arcsin-based probability-to-angle mapping as Eq.~\ref{eq:cry_theta}, but with the resulting angle clipped to a small maximum $\Theta_{\mathrm{cap}}$ ($=\pi/6$ in this work). This cap is what keeps the term a light, parametrised correction rather than a competing transfer channel: even for a near-zero drive, where the raw arcsin mapping could otherwise assign an appreciable rotation, $\vartheta^{YY}_{ij}$ is bounded well below the angles typically assigned to the dominant, strongly-directed $\operatorname{CRY}$ pairs. Pairs whose drive exceeds $\tau_{YY}$ instead receive the full bidirectional $\operatorname{CRY}$ layer of Eq.~\ref{eq:cry_relation}, so the two gate types partition the scheduled pairs rather than acting jointly on the same pair.

The choice of $\hat{Y}_i\hat{Y}_j$ over other two-qubit generators is deliberate: because $\hat{Y}\ket{0}\propto\ket{1}$ and $\hat{Y}\ket{1}\propto\ket{0}$, the operator only ever flips both qubits together, so $U^{YY}_{ij}$ conserves the total excitation parity $\prod_r\hat{Z}_r$ of the pair -- it can jointly excite, de-excite, or exchange excitation between $i$ and $j$, but cannot alter the parity of the pair in isolation. This is the natural symmetry for a bath-mediated fluctuation between structurally proximate but energetically balanced residues, in contrast to $\operatorname{CRY}$, which explicitly breaks this symmetry to encode a preferred direction of transfer. In this sense, $\operatorname{CRY}$ and $\operatorname{IsingYY}$ act as complementary entangling primitives on the same schedule: the former imprints the directed, energetically biased component of residue-residue coupling for pairs with a resolvable excitation gradient, while the latter adds a small, parity-conserving correction representing residual, direction-less thermal exchange for pairs without one.

\subsection{Measurement and empirical ensemble generation}
\label{sec:measurement}

After the final interaction layer, the circuit was measured on the $Z$-basis, each measurement shot returned a length-\(N\) Pauli-\(Z\) sample bitstring. We record $Z_r=+1$ for outcome $|0\rangle$ (ground state) and $Z_r=-1$ for outcome $|1\rangle$ (excited state). Each measurement yields a single microstate vector $\mathbf{Z}=(Z_1,Z_2,...,  Z_r,\dots,Z_N)$. Repeated independent executions of the circuit produce an empirical ensemble of such solvation states of AAs. The frequency count of a particular microstate appearing in this ensemble statistics, reflect the depth of the corresponding energy basin in the energy landscape. Deeply stabilised configurations are sampled with high probability, whereas high-energy configurations with disrupted contacts are sampled rarely, see fig.~\ref{fig:trp_energy_funnel}.

Each measured bit-string represents one residue-level microstate of the protein. The total energy assigned to a sampled microstate was computed as a diagonal residue-wise Hamiltonian,

\begin{align} 
    \hat{\mathcal{H}}_{\mathrm{tot}} = \sum_{r=1}^{N} \left[ E_{0,r}\ket{0_r}\bra{0_r} + E_{1,r}\ket{1_r}\bra{1_r} \right], \label{eq:Htot_operator} 
\end{align} 

where $E_{0,r}$ and $E_{1,r}$ are the ground- and excited-state energy contributions assigned to residue $r$, respectively. For a sampled configuration, this gives the scalar Gibbs free energy as

\begin{align} 
    G_{model}(Z^{(m)}) = \sum_{r=1}^{N} \left[ \mathcal{I}\!\left(Z_r^{(m)}=+1\right)E_{0,r} + \mathcal{I}\!\left(Z_r^{(m)}=-1\right)E_{1,r} \right], \label{eq:Etot} 
\end{align} 

where $m$ indexes the sampled microstate, $N$ is the number of residues, and $\mathcal{I}(\cdot)$ is an indicator function. This Hamiltonian is diagonal in the measurement basis, so the energy of a sampled state is obtained by summing the residue-level energy assigned to the observed state of each residue. The interaction layers affect the ensemble by changing the probability of sampled microstates, whereas the Hamiltonian provides a fixed scoring function for assigning an energy to each sampled configuration. Repeating Eq.~\ref{eq:Etot} over all sampled microstates gives the empirical ensemble energy distribution, 

\begin{align}
    P(G_{model}) = \frac{1}{M} \sum_{m=1}^{M} \delta\!\left(G_{model}-G_{model}(Z^{(m)})\right),
    \label{eq:energy_distribution}
\end{align}
which, in the large-$M$ limit, converges to the histogram shown in Fig.~\ref{fig:trp_energy_funnel} and constitutes the empirical density of states sampled by the circuit. Here, a diagonal Hamiltonian may appear to neglect residue-residue cooperativity, since it contains no explicit coupling term of the form $Z_rZ_{r'}$. Writing $a_r = (E_{0,r}+E_{1,r})/2$ and $b_r = (E_{0,r}-E_{1,r})/2$, Eq.~\ref{eq:Htot_operator} can equivalently be expressed in the Pauli basis as $\hat{\mathcal{H}}_{\mathrm{tot}} = \sum_r \left(a_r\hat{I}_r + b_r\hat{Z}_r\right)$, confirming that it is an explicitly non-interacting, field-only operator: every computational basis state $\ket{Z^{(m)}}$ is one of its eigenstates, so $G_{model}(Z^{(m)}) = \bra{Z^{(m)}}\hat{\mathcal{H}}_{\mathrm{tot}}\ket{Z^{(m)}}$ is assigned exactly, with no ambiguity between the measurement basis and the scoring rule. The absence of coupling terms is nonetheless not a limitation, because cooperativity does not need to enter through $\hat{\mathcal{H}}_{\mathrm{tot}}$ at all. The Born rule gives the sampling probability as $P(Z^{(m)}) = |\langle Z^{(m)}|\psi\rangle|^2$, where $\ket{\psi}$ is the state produced by the entangling interaction layers. Since $\ket{\psi}$ is entangled, it cannot be written as a product state $\bigotimes_r\ket{\phi_r}$, and consequently

\begin{align}
    P(Z_1,\dots,Z_N) \neq \prod_{r=1}^N P(Z_r),
\end{align}

so residue outcomes are statistically correlated at the level of the sampled distribution, even though $\hat{\mathcal{H}}_{\mathrm{tot}}$ itself contains no coupling terms. The funnel-like concentration of probability onto low-energy, contact-rich configurations in Fig.~\ref{fig:trp_energy_funnel} is therefore generated entirely upstream, by the interaction layers reshaping $P(Z^{(m)})$ so that cooperatively stabilised configurations are sampled more often while disrupted ones are suppressed -- not by $\hat{\mathcal{H}}_{\mathrm{tot}}$, which only scores whatever configuration is handed to it. Adding an explicit $\sum_{r<r'}J_{rr'}Z_rZ_{r'}$ term would therefore double-count correlations already present in the sampled ensemble rather than adding new physics. The two components remain complementary: repeated measurement converts the correlated distribution imprinted by the interaction layers into the empirical ensemble of Eq.~\ref{eq:energy_distribution}, and in the large-$M$ limit the sample mean of $G_{model}$ converges to the quantum expectation value $\mathrm{Tr}(\rho\,\hat{\mathcal{H}}_{\mathrm{tot}})$ -- consistent with treating the circuit output as an effective, residue-correlated ensemble built on a strictly local, two-state-per-residue scoring rule.

\subsection{Residue-level energetic sensitivity}
\label{sec:energy_sensitivity}

The influence of an individual residue state on the coarse-grained Gibbs-energy distribution was quantified using a conditional Kullback--Leibler divergence:
\begin{align}
    \mathfrak{I}_{\sigma}(r)
    &=
    D_{\mathrm{KL}}
    \left[
    P
    \left(
    G_{model}(Z^{(m)})
    \mid
    Z_r=\sigma
    \right)
    ,\middle|,
    G_{model}(Z^{(m)})
    \right],
    \qquad
    \sigma\in{+1,-1}.
    \label{eq:energetic_sensitivity}
\end{align}

A large value of ($\mathfrak{I}_{-1}(r)$) indicates that conditioning on excitation of residue ($r$) substantially changes the global energy distribution.

\subsection{Directional excitation information}
\label{sec:DEI}

To quantify directional statistical dependence between residue excitation states, the binary excitation variable was defined as
\begin{align}
    X_r^{(m)}
    &=
    \mathcal{I}
    \left(
    Z_r^{(m)}=-1
    \right).
\end{align}
For an ordered residue pair ($j\rightarrow i$), the directional excitation information score was calculated as
\begin{align}
    D_{j\rightarrow i}
    &=
    \widehat{P}
    \left(
    X_i=1
    \mid
    X_j=1
    \right)
    \ln
    \left[
    \frac{
    \widehat{P}
    \left(
    X_i=1
    \mid
    X_j=1
    \right)
    +
    \epsilon
    }{
    \widehat{P}
    \left(
    X_i=1
    \right)
    +
    \epsilon
    }
    \right].
    \label{eq:dei}
\end{align}
This score measures whether excitation of residue ($j$) increases or decreases the observed excitation probability of residue ($i$) relative to its marginal probability. Because conditioning is directional, ($D_{j\rightarrow i}$) is not generally equal to ($D_{i\rightarrow j}$). It is therefore distinct from mutual information, which is symmetric by definition. When a symmetric residue-pair affinity was required for clustering or network visualisation, it was calculated as
\begin{align}
    A_{ij}
    &=
    \left|
    C_{i\rightarrow j}
    \right|
    +
    \left|
    C_{j\rightarrow i}
    \right|.
\end{align}

\subsection{Implementation and reproducibility}
\label{sec:comp_reprod}

Structure processing and residue-level geometric analysis were performed using MDAnalysis \cite{MichaudAgrawal2011,Gowers2016}. SASA was calculated using FreeSASA \cite{Mitternacht2016FreeSASA}, and the circuits were constructed and executed using PennyLane \cite{bergholm2022}. The Circuit simulation used \texttt{lightning.gpu} backend, compatibile with GPUs. The default measurement shots were $M=10^6$ to ensure convergence of patterns, however the number of shots required might be less for a real quantum computer as it may be able to justify the randomness naturally, which can drop significantly to $10^4$ to generate a similar pattern. Furthermore, all the parameters for the circuit---backend, number of interaction layers, shot count, transfer-scale parameter, numerical regularisation constants, edge cutoffs, structural weights, temperature, chain selection, and random seeds were fixed before the final runs. Configuration files, input structures, source code, random seeds, and the scripts used to generate the reported figures are provided in the accompanying github repository and Supplementary Information (see appex.~\ref{app:params}). The code can be used as a python package, additionally docs for code implementation and its usage are provided in the github repository.

\section{Results}
\label{sec:results}

Herein, we report results for two complementary Trp-cage-based mini-protein systems: the canonical Trp-cage construct TC5b (PDB 1L2Y) and the disulfide-cyclized Trp-cage-fortified exenatide chimera (PDB 9GDL), analysed using the quantum microstate ensemble framework described in Sec.~\ref{sec:methods}. TC5b is a 20-residue designed miniprotein with sequence NLYIQWLKDGGPSSGRPPPS and is widely used as a benchmark for protein-folding and energy-landscape studies because it forms a compact, protein-like structure despite its small size \cite{Neidigh2002TrpCage,Marinelli2009TrpCage}. Its native fold consists of an N-terminal $\alpha$-helix, a short $3_{10}$-helical region, and a C-terminal polyproline segment arranged around the central Trp6 residue to form the characteristic hydrophobic cage \cite{Neidigh2002TrpCage,Lee2013TrpCage}. Hydrophobic packing around Trp6, together with tertiary interactions such as the Asp9--Arg16 salt bridge, provides a well-characterised minimal interaction network for evaluating whether the framework identifies structurally important residues and correlated interactions within a compact fold. In contrast, 9GDL is a 25-residue Trp-cage-fortified exenatide derivative with sequence EEECVRLYIQWLKDGGPSSGRPPPC, in which the terminal cysteine residues form a disulfide-cyclized construct \cite{Horvath2024TrpCageExenatide}. The system retains the Trp-cage motif within a functionally relevant GLP-1 receptor agonist context, where cage compactness has been associated with thermal stability, aggregation behaviour, insulinotropic activity, and receptor binding \cite{Horvath2024TrpCageExenatide}. Relative to TC5b, 9GDL introduces sequence extension, additional charged residues, and a covalent topological constraint, providing a more complex test of the framework's sensitivity to chemical stabilization and long-range interaction-network organization. Together, the two systems enable evaluation of the model across a minimal, well-characterised folding motif and a chemically constrained, functionally relevant derivative. The final circuit-sampled ensembles were analysed using residue ground-state occupancy, relative configuration-energy distributions, residue-wise energetic sensitivity, mutual information, directional excitation information, and dimer/trimer-level affinity.

\subsection{Effect of entanglement block}
\label{subsec:population_redistribution}

\begin{figure}[ht]
    \centering
    \vspace{-0.8em}
    \includegraphics[width=1.0\textwidth]{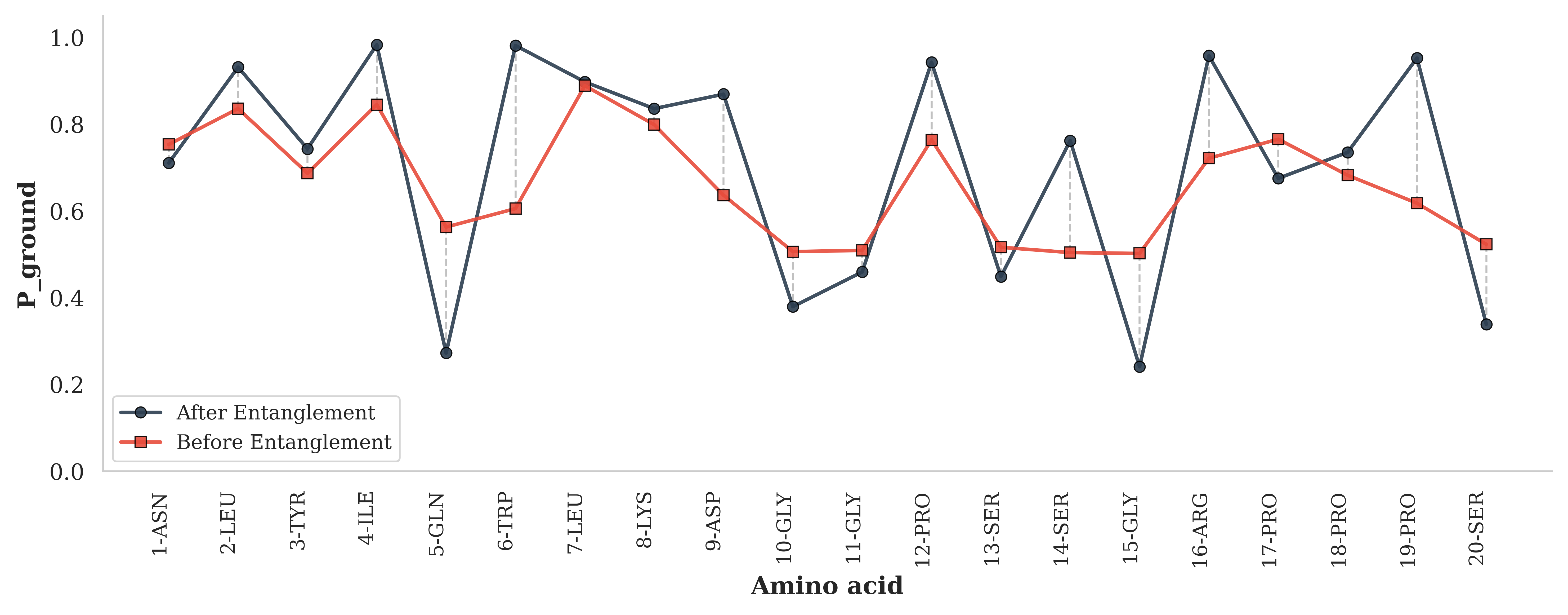}
    \vspace{0.5em}
    \includegraphics[width=1.0\textwidth]{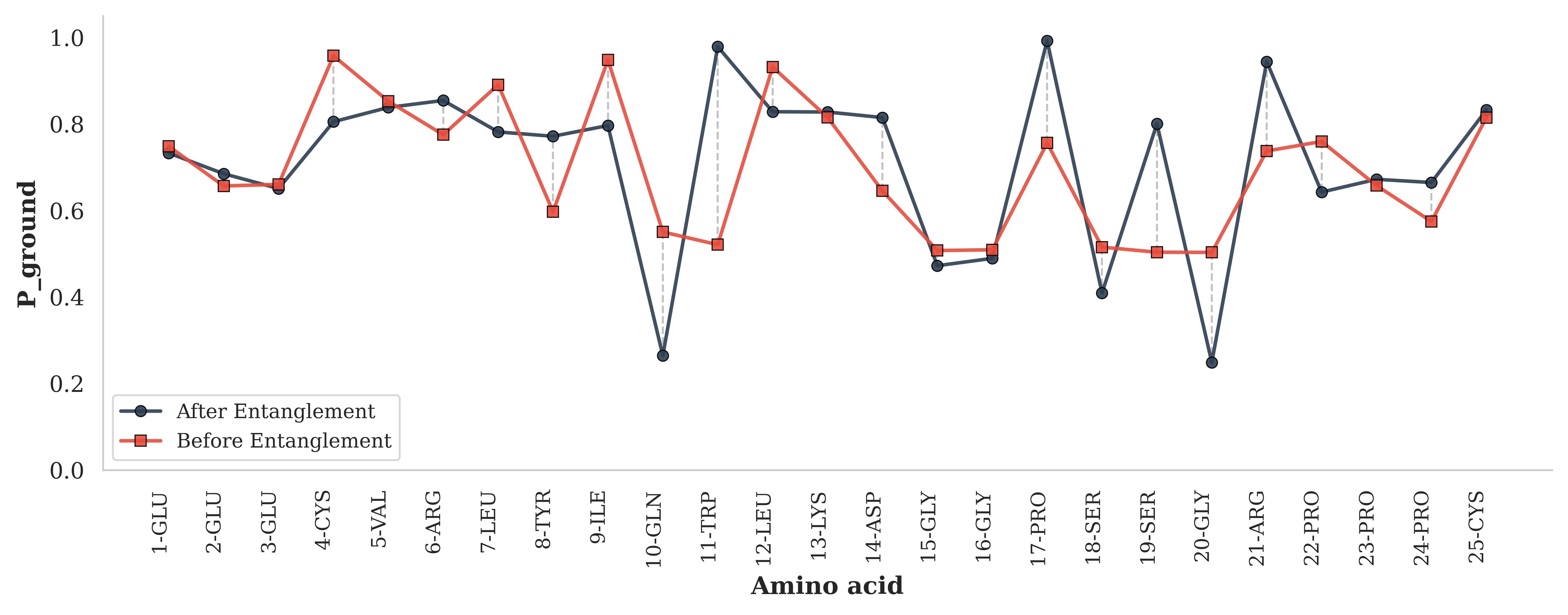}
    \caption{
    \textbf{Ground state comparison of before and after entanglement for PDB 1L2Y and PDB 9GDL}
    The plots compare residue-wise ground-state probabilities before and after application of the structure-derived entangling circuit. Deviations from the pre-entanglement profile show that the circuit redistributes residue-state probabilities.}
    \label{fig:ground_comparison}
\end{figure}

We first examined whether the entanglement block modifies the independent residue-state distribution established during single-qubit initialisation. Figure~\ref{fig:ground_comparison} compares the residue-wise ground-state probabilities before and after application of the structure-derived interaction layers. In both 1L2Y and 9GDL, entanglement produced heterogeneous changes across the sequence rather than a uniform increase or decrease in ground-state occupancy. For 1L2Y, marked redistribution occurred around residues associated with the hydrophobic core, including Ile4 and Trp6, as well as within the Gly/Pro/Ser-rich region. For 9GDL, pronounced changes were observed around Trp11 and the Gly16--Pro17--Ser18--Ser19 segment, together with shifts at several charged and terminally constrained residues. The post-entanglement profiles therefore differed systematically from the independent-residue baselines. These results show that the entanglement block transforms the initial solvation-derived residue propensities into a structure-conditioned thermodynamic ensemble. The non-uniform population shifts indicate that the final residue-state occupancies depend on the topology and strength of the encoded interaction network, rather than on intrinsic residue solvation energetics alone. Moreover, in 9GDL, the different responses of Cys4 and Cys25 further show that the disulfide-linked residues are not assigned a uniform stabilising shift. Instead, their occupancies reflect their distinct local environments and their coupling to the wider residue-interaction network.

\subsection{Energy distribution}
\label{subsec:energy_distributions}

\begin{figure}[ht]
    \centering
    \includegraphics[width=0.9\textwidth]{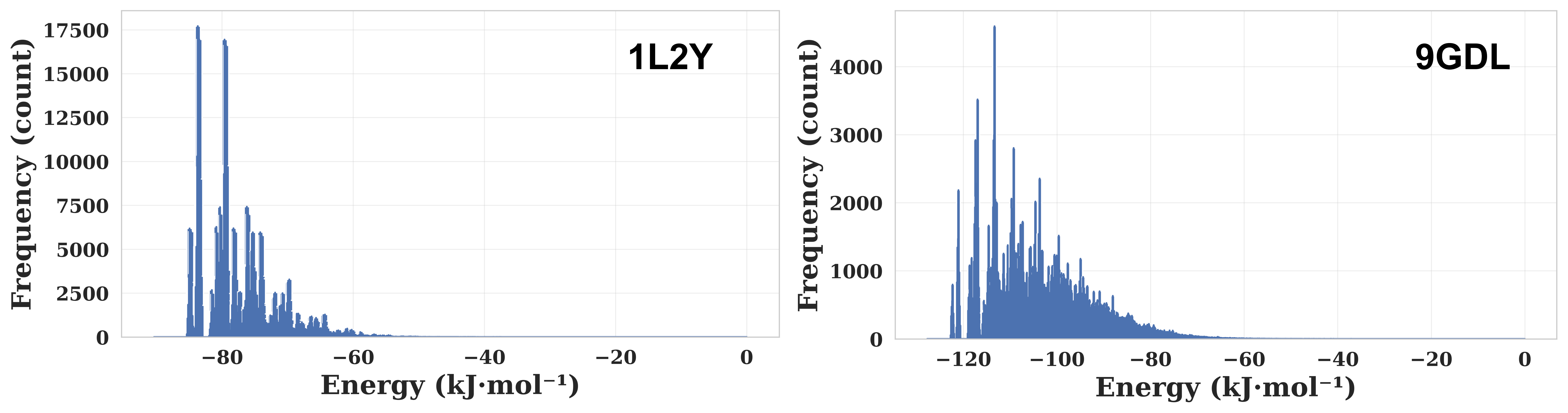}
    \caption{
    \textbf{Relative sampled energy distributions for 1L2Y and 9GDL.}
    Each histogram was constructed from $10^6$ circuit-sampled residue-state microstates. The (x)-axis shows the configuration Gibbs free energy, discretised into $10^5$ bins, while the (y)-axis reports the number of sampled microstates per bin. Even though the two histograms use different bin widths, their overall distributional profiles remain qualitatively informative. Under the same scoring convention, 9GDL samples a broader and lower-scored energy regime than 1L2Y.
    }
    \label{fig:energy_distributions}
\end{figure}

The sampled coarse-grained Gibbs-energy distribution was used to examine how the circuit-generated residue-state microstates were organised under the fixed scoring Hamiltonian (see~\ref{eq:Etot}). Figure~\ref{fig:energy_distributions} shows that 1L2Y is concentrated within a relatively narrow energy range, with several sharp peaks and a declining high-energy tail. In contrast, 9GDL exhibits a broader distribution with a wider spread of sampled energies, though the ensembles are lower energies than 1L2Y. The multiple low-energy peaks represent distinct residue-state configurations assigned similar favourable scores by the effective Hamiltonian. The intermediate and high-energy regions contain configurations with progressively larger contributions from solvation-excited residue states.

\subsection{Energetic sensitivity}
\label{subsec:energetic_sensitivity}

\begin{figure}[ht]
    \centering
    \includegraphics[width=0.48\textwidth]{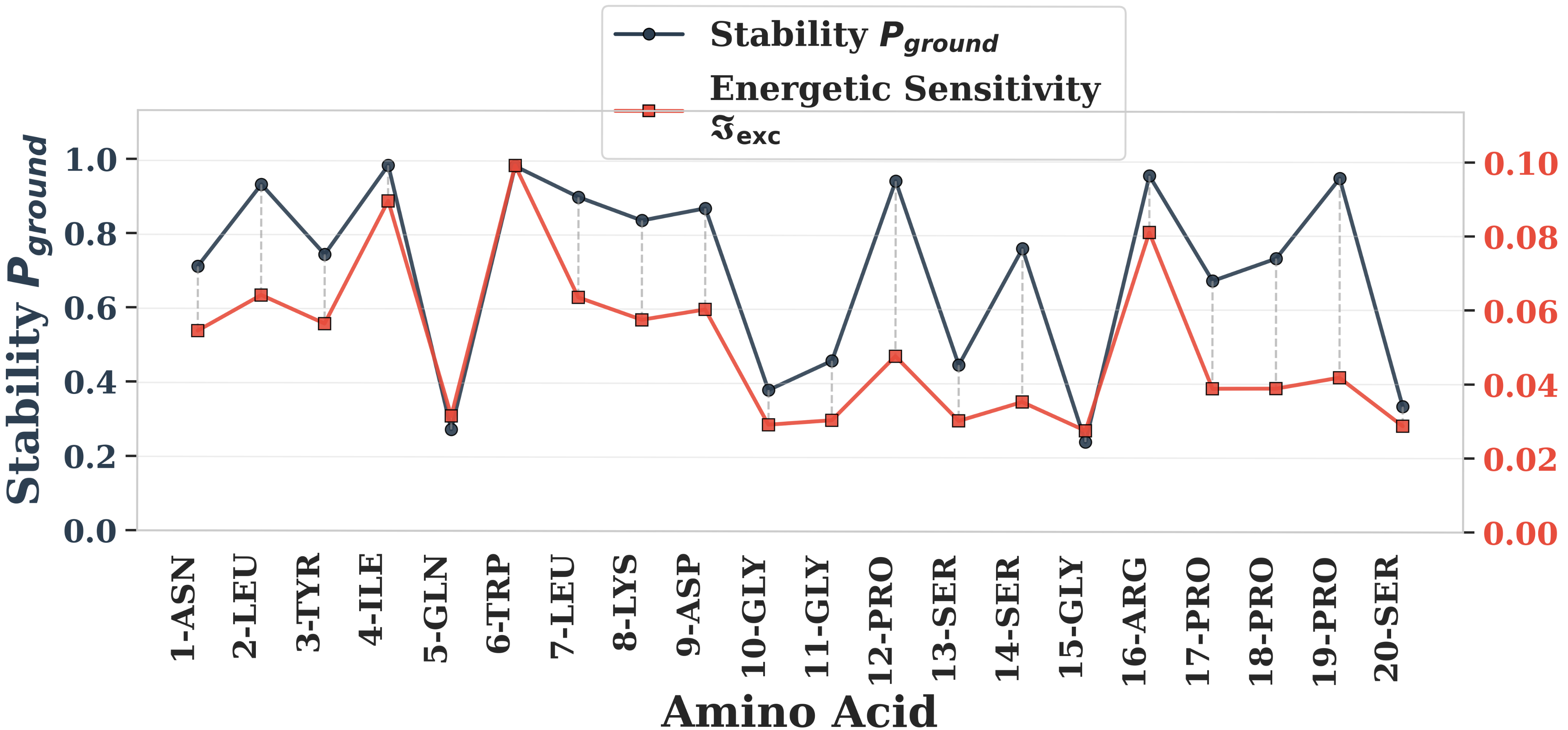}
    \includegraphics[width=0.48\textwidth]{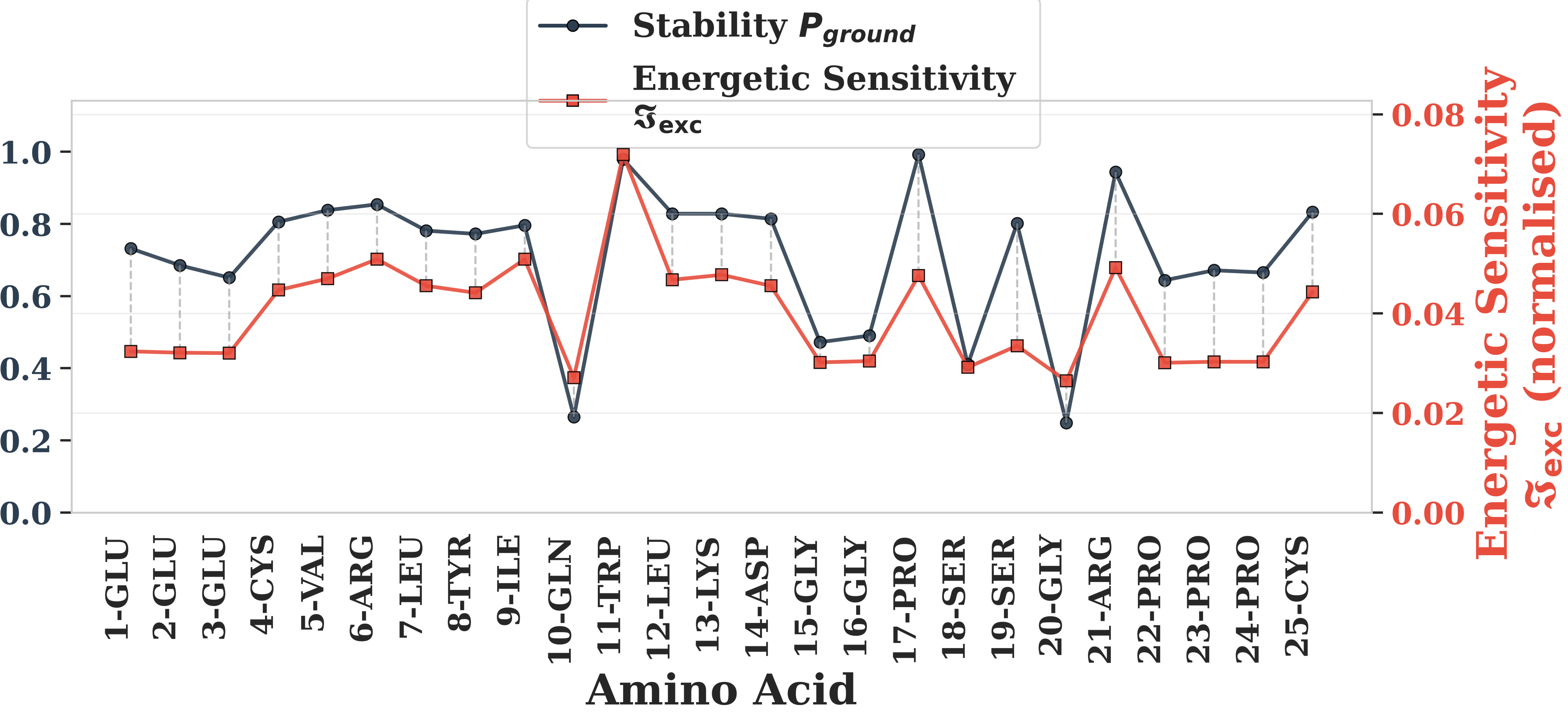}
    \includegraphics[width=0.48\textwidth]{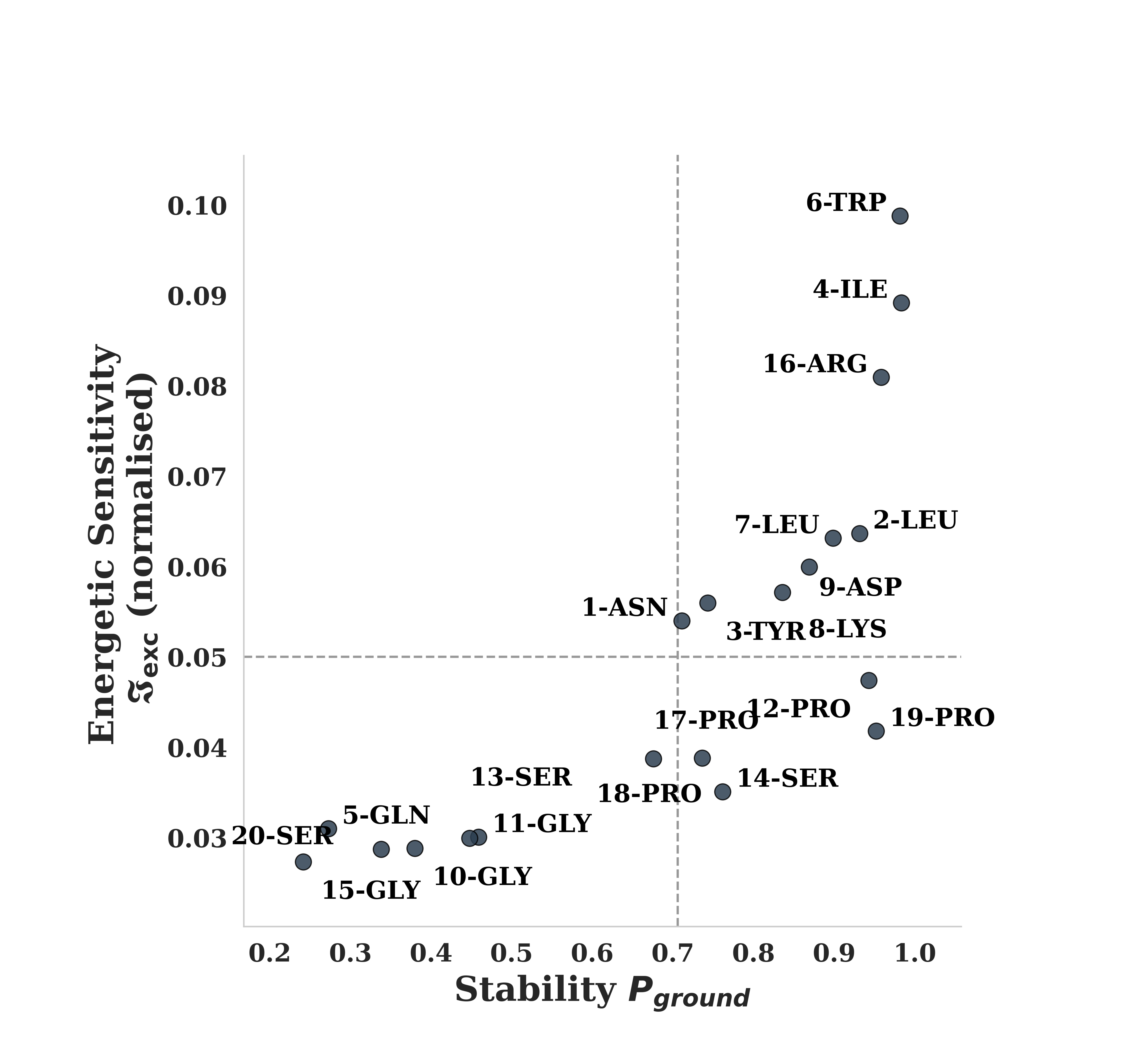}
    \includegraphics[width=0.48\textwidth]{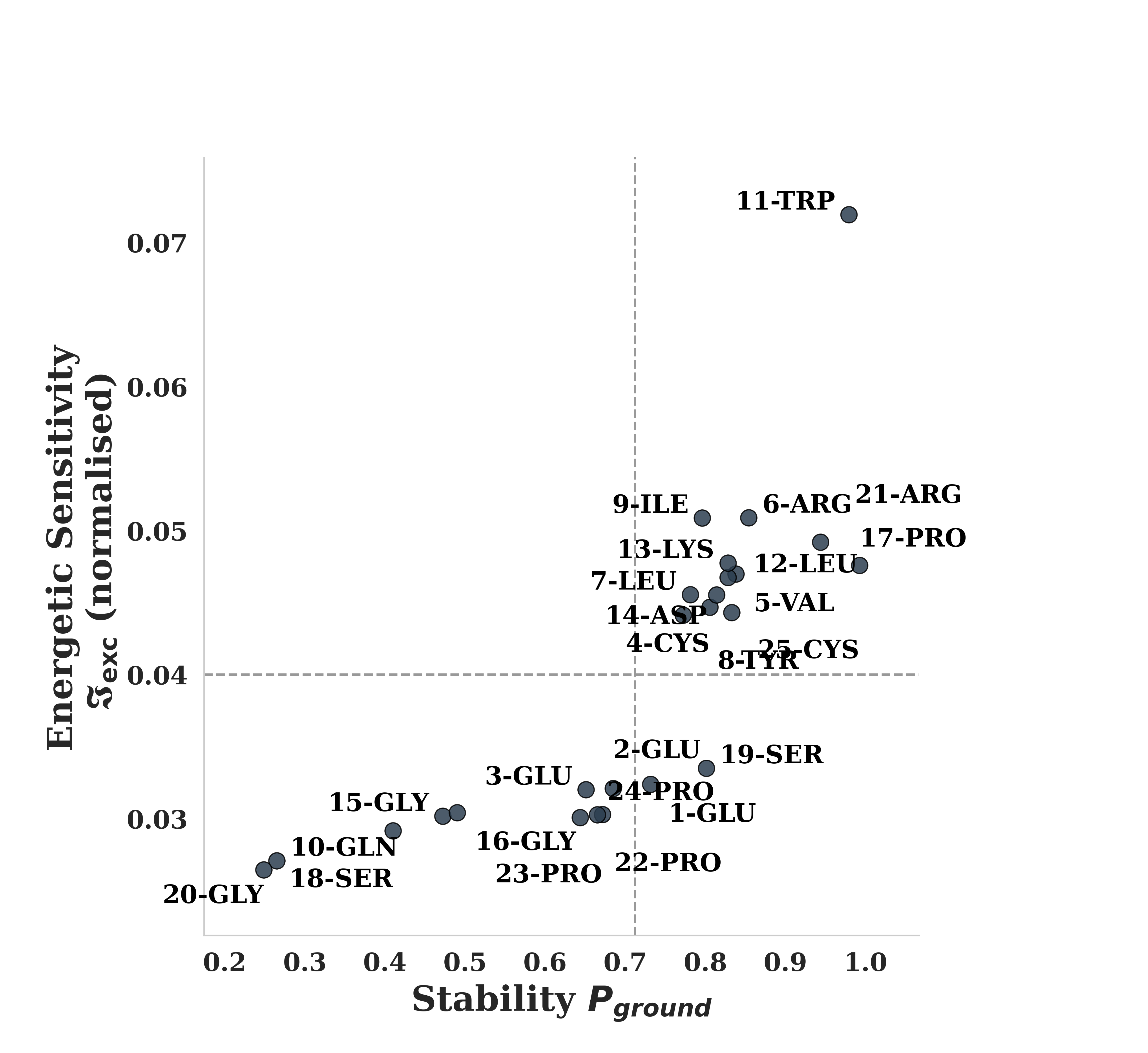}
    \caption{
    \textbf{Residue ground-state stability and energetic sensitivity.}
    The top panels show residue-wise profiles, while the bottom panels compare the two quantities directly. While the left side is for 1L2Y and the right side is for 9GDL structures respectively. Ground-state occupancy measures the persistence of the stabilised residue state, whereas energetic sensitivity quantifies the change in the global energy distribution when that residue is conditioned to be excited.}
    \label{fig:stability_sensitivity}
\end{figure}

To further understand the influence of each residue, we calculated energetic sensitivity as defined in sec~\ref{sec:energy_sensitivity}. We further compared it with the residue stability (ground-state occupancy) in Figure~\ref{fig:stability_sensitivity}. Residue ground-state occupancy and excited-state energetic sensitivity provide complementary information. The two quantities were related but not equivalent: the ground-occupancy describes local state stability, whereas sensitivity measures a residue's influence on the sampled ensemble.

In 1L2Y, energetic sensitivity was concentrated around the N-terminal hydrophobic and aromatic core. Trp6 was the dominant sensitivity site, with additional contributions from Ile4, Tyr3, Leu7, Asn1, Leu2, and Arg16. In contrast, several residues within the Gly/Ser/Pro-rich region showed greater local lability but lower influence on the global energy distribution. The model therefore separated a Trp6-centred energetic-control region from a more locally labile central and C-terminal segment.

A corresponding pattern was observed in 9GDL, where Trp11 was the dominant sensitivity site, supported by Tyr8, Ile9, Leu12, Arg6, Lys13, and Arg21. Although the Cys4--Cys25 disulfide bond contributes to the interaction network, the strongest energetic control remained centred on the Trp-containing hydrophobic core. Compared with 1L2Y, sensitivity in 9GDL can be seen distributed across a larger set of residues, indicating a more dispersed organisation of energetic influence arising from the extended sequence, cyclisation, and altered contact network.

\subsection{Pairwise information analysis identifies construct-specific communication modules}
\label{subsec:pairwise_coupling}

\begin{figure}[ht]
    \centering
    \includegraphics[width=0.48\textwidth]{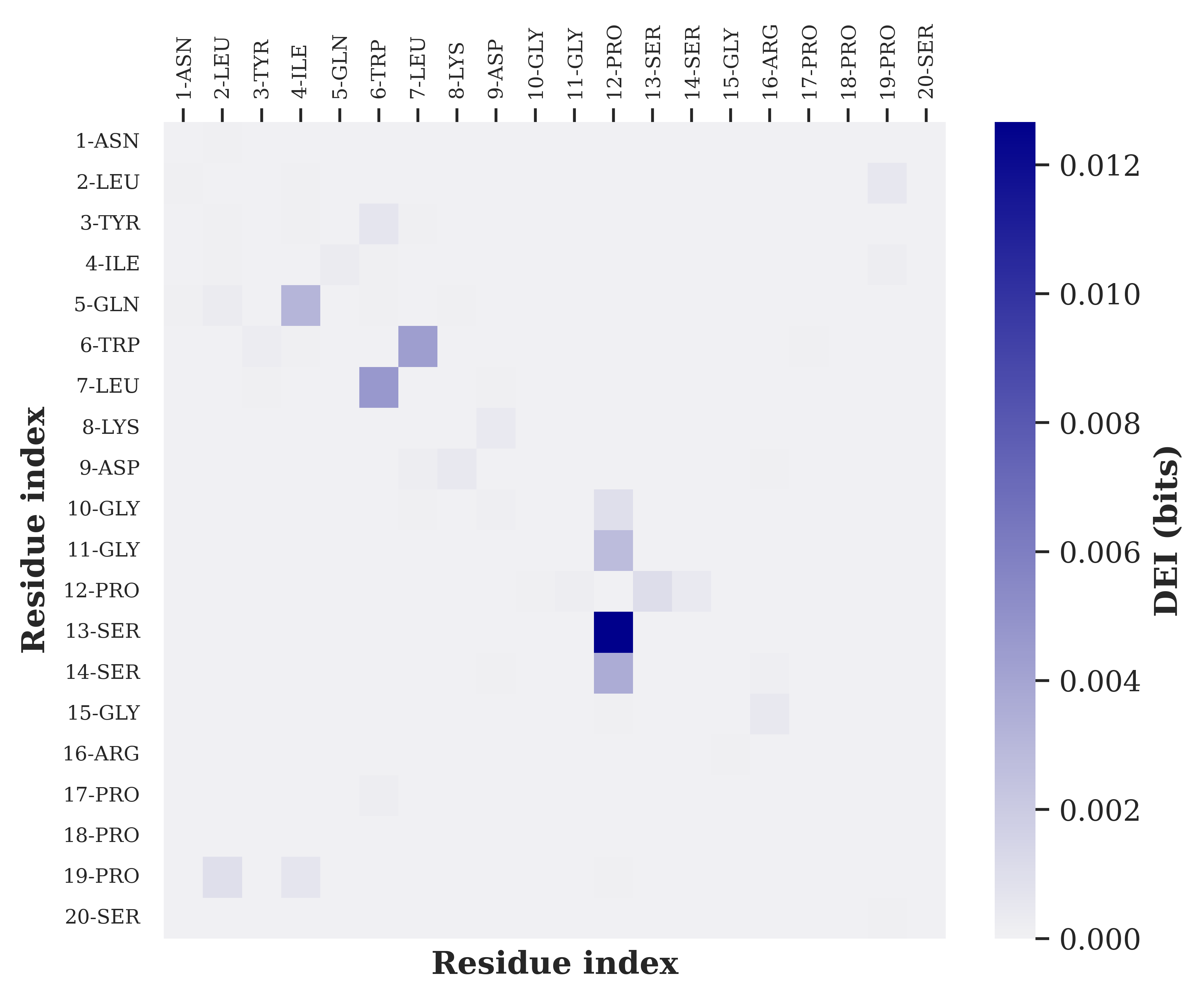}
    \includegraphics[width=0.48\textwidth]{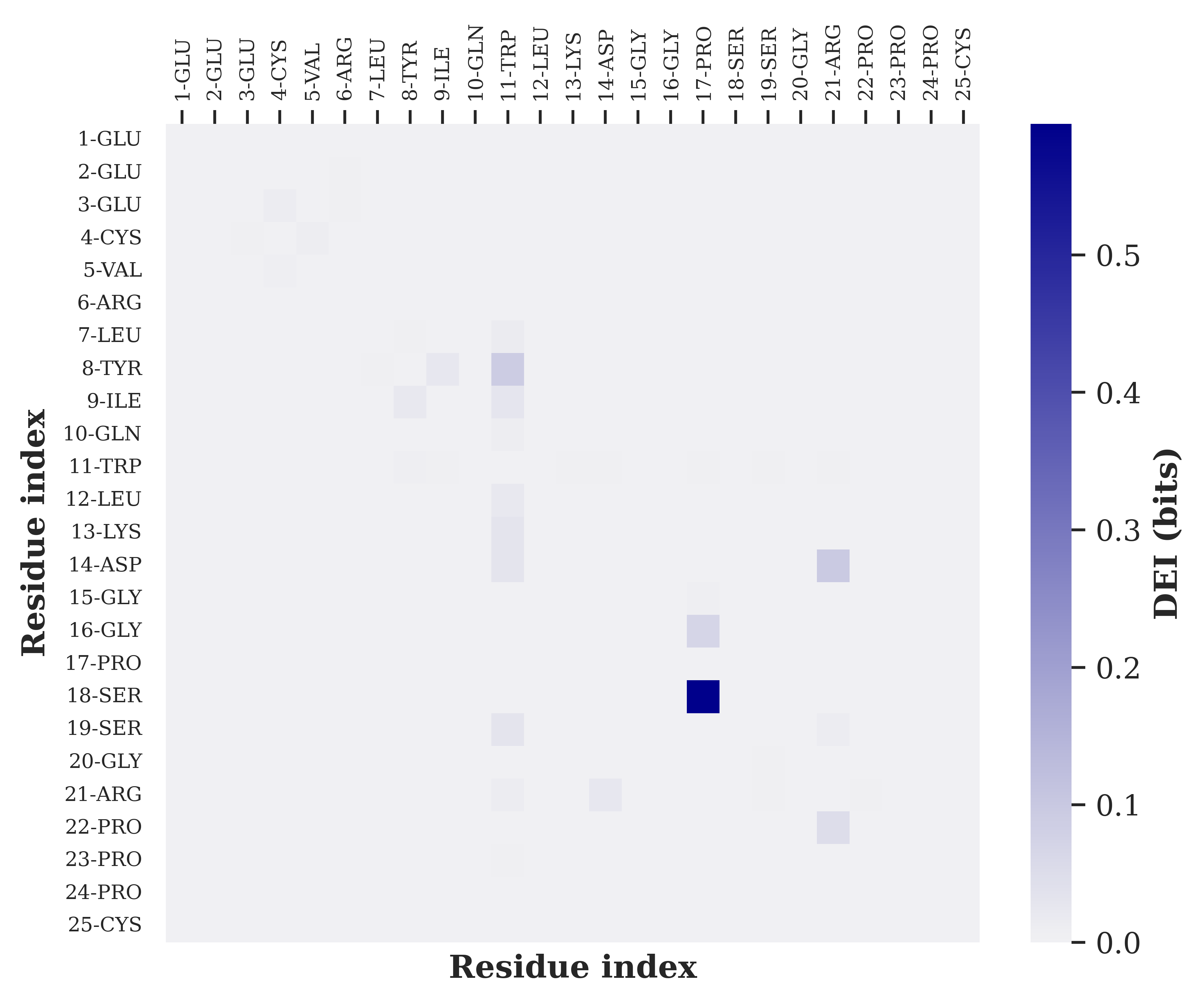}
    \caption{
    \textbf{Directional and symmetric residue-state coupling.}
    Left: Directional Excitation-Information (DEI) map for 1L2Y, Right: DEI map for 9GDL. A directional signal \(D_{j\rightarrow i}\) reports whether excitation of source residue \(j\) (where $j$ is a column and $i$ is a row) changes the excitation statistics of target residue \(i\).
    }
    \label{fig:directional_and_mi}
\end{figure}

Pairwise residue-state dependence was analysed using directional excitation information (DEI) (Fig.~\ref{fig:directional_and_mi}). Directional excitation information identifies ordered source-target statistical dependence between residue states. The DEI was calculated as defined in sec.~\ref{sec:DEI}. In 1L2Y, the strongest directional signals were concentrated within the Gly11--Pro12--Ser13--Ser14 segment, with Pro12 acting as a principal conditioning source for Ser13 and Ser14. The mutual-information analysis also highlighted Pro12--Ser13 and Gly11--Pro12, together with Trp6--Leu7 and Ile4--Gln5. These results separated two residue-level roles: the Trp6-centred core dominated energetic sensitivity, whereas the Gly11--Pro12--Ser13--Ser14 segment dominated local directional communication.

The corresponding communication module in 9GDL was shifted to the Gly16--Pro17--Ser18--Ser19 region. The strongest directional effect was associated with Pro17 as a conditioning source for Ser18. Pairwise signals involving Tyr8--Ile9 and Trp11--Leu12 were also retained, linking the directional module to the Trp11-centred energetic core. This separation is important. The model does not collapse all high-ranking pairs into one generic allostery score. It distinguishes energetic control, symmetric pair dependence, and directional excitation transfer.

\subsection{Motif-level Couplings}
\label{sec:results-motif-allostery}

It can be noted from the DEI in fig.~\ref{fig:directional_and_mi}, it is highly sparse, that indicates, no one residue dictates the coupling, rather the solvation of a residue is a function of more than one residues. So the residue-level directional effects were next aggregated over overlapping peptide dimers, consecutive peptide trimers, and peptide/non-covalent triads. The resulting affinity matrices were interpreted as relative within-construct communication maps because the motifs overlap and the number of available motifs differs between the two systems (1L2Y and 9GDL).

\begin{figure}[ht]
    \centering
    \includegraphics[width=0.48\textwidth]{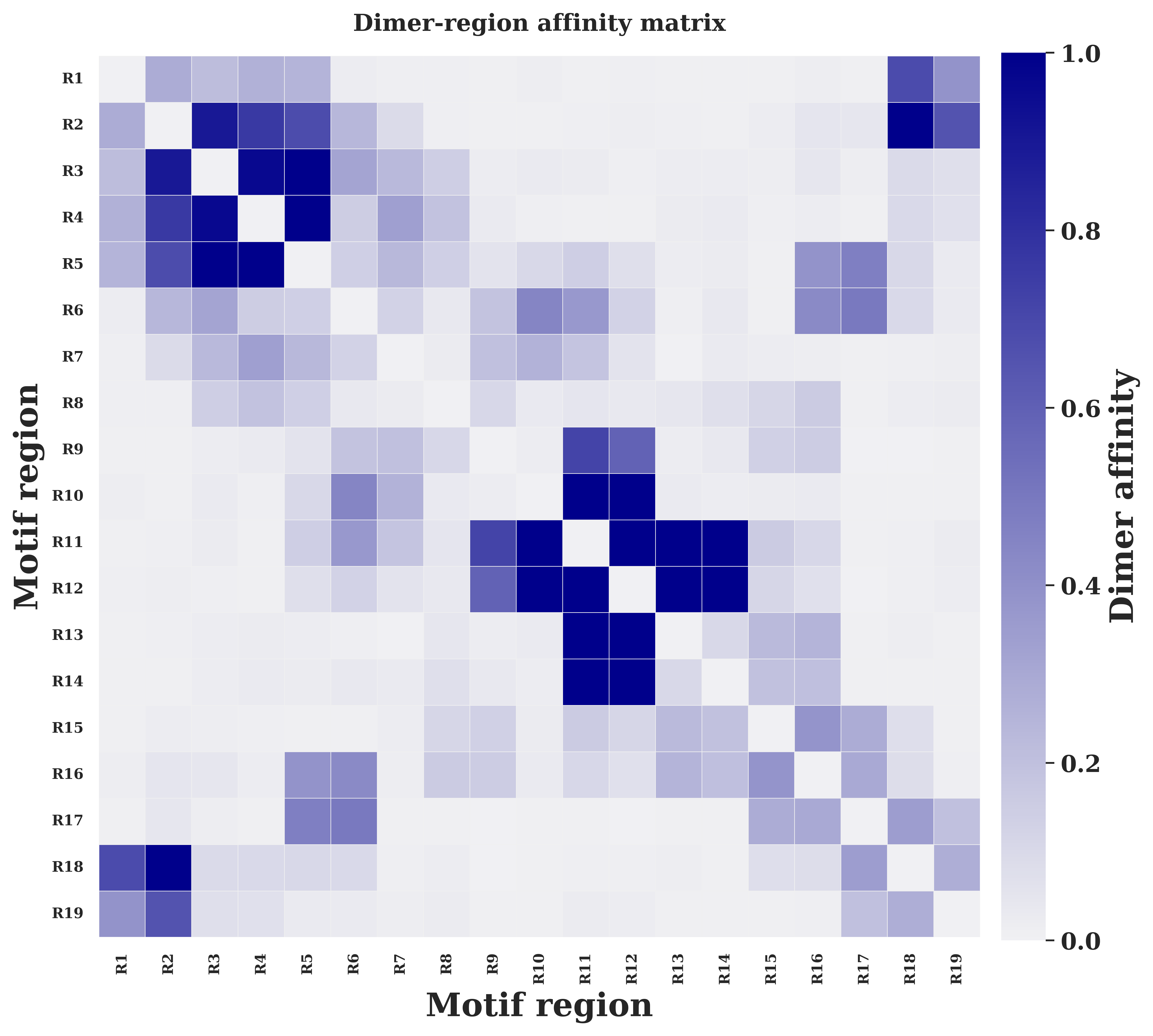}
    \includegraphics[width=0.48\textwidth]{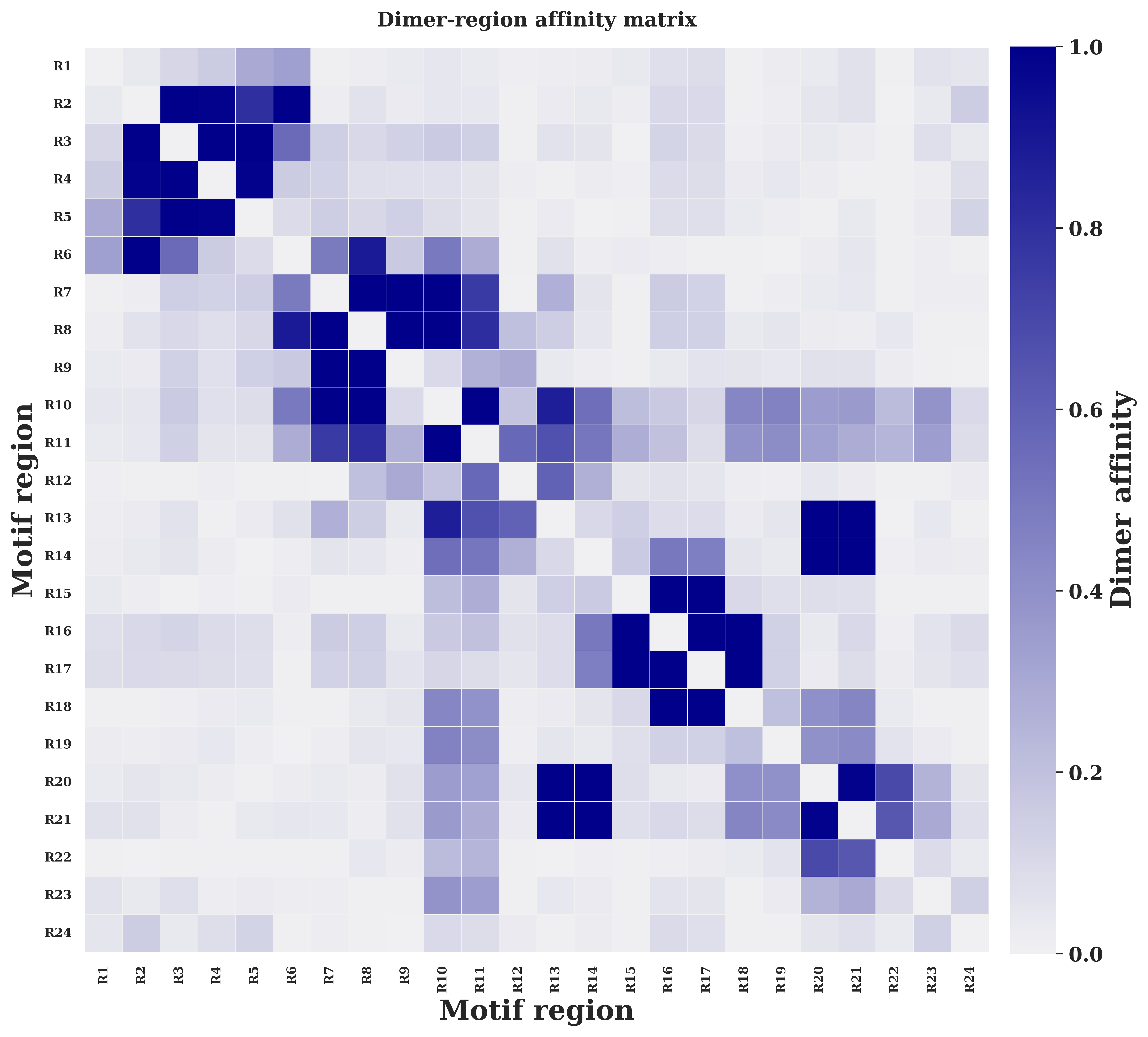}
    \caption{
    \textbf{Dimer motif affinity maps for 1L2Y and 9GDL.}
        The left and right panels show the maps for 1L2Y and 9GDL, respectively. Each rank $Ri$ denotes a consecutive peptide dimer; for example, $R1$ represents Asn1--Leu2 and $R2$ represents Leu2--Tyr3.
    }
    \label{fig:dimer_affinity}
\end{figure}

In 1L2Y, the dimer-level map preferentially connected Gly/Pro/Ser-rich motifs with motifs containing Gln and Trp (Figure~\ref{fig:dimer_affinity}). This pattern was consistent with communication between the flexible central segment and the Trp6-associated core. The corresponding trimer and triad maps retained this organisation, with strong links between Gln/Trp/Pro-containing motifs and Asp/Gly/Ser-containing motifs.

In 9GDL, the dominant motif-level links shifted towards acidic and cysteine-containing terminal motifs and the central Arg/Leu/Trp-containing region. The trimer and triad maps similarly connected Glu/Cys/Ile-containing motifs to the central hydrophobic and charged region. The 9GDL network therefore reflected the combined effects of terminal constraints, central hydrophobic packing, and the shifted Pro/Gly/Ser communication segment.

The agreement between residue-level and motif-level analyses indicates that the strongest directional signals were not isolated pairwise events. They persisted when residue effects were aggregated over structurally supported local groups. Further methodological details and the trimer affinity maps are provided in Appendix~\ref{app:motifs} and Figures~\ref{fig:app_trimer_affinity_1l2y} and~\ref{fig:app_trimer_affinity_9gdl}.

\section{Discussion}
\label{sec:discussion}

\begin{figure}[ht]
    \includegraphics[width=\linewidth]
    {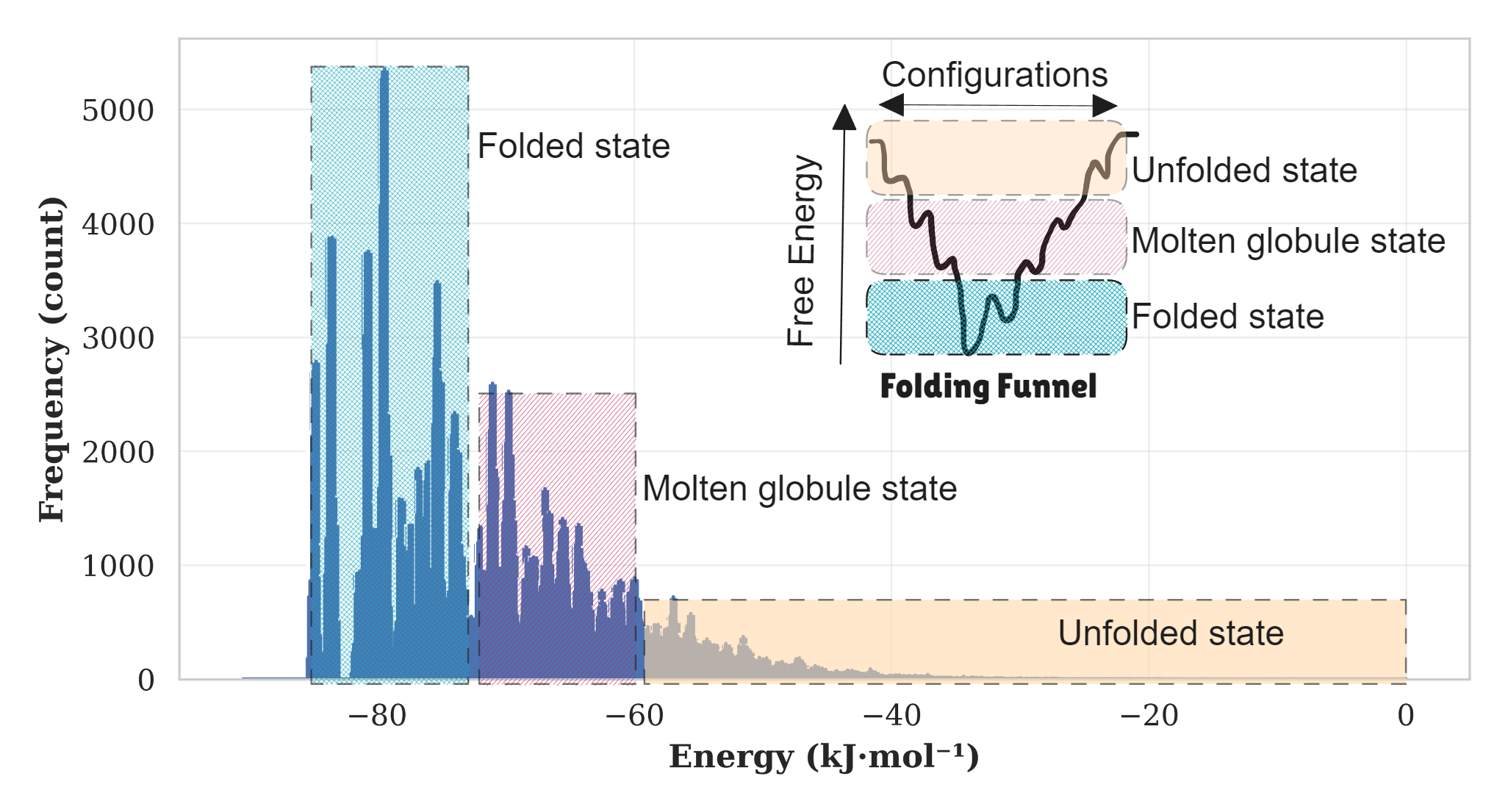}
    \caption{
    \textbf{Circuit-sampled energy distribution and folding-funnel interpretation for Trp-cage (PDB:1L2Y).}
        The histogram shows the configuration energies of $10^{6}$ sampled microstates using $10^{5}$ bins. The distribution contains multiple low-energy peaks, an intermediate-energy shoulder, and a diffuse high-energy tail. The inset provides a qualitative funnel representation of these relative-energy regimes and their underlying landscape ruggedness.
    }
    \label{fig:trp_energy_funnel}
\end{figure}

\begin{figure}[ht]
    \includegraphics[width=0.8\linewidth]
    {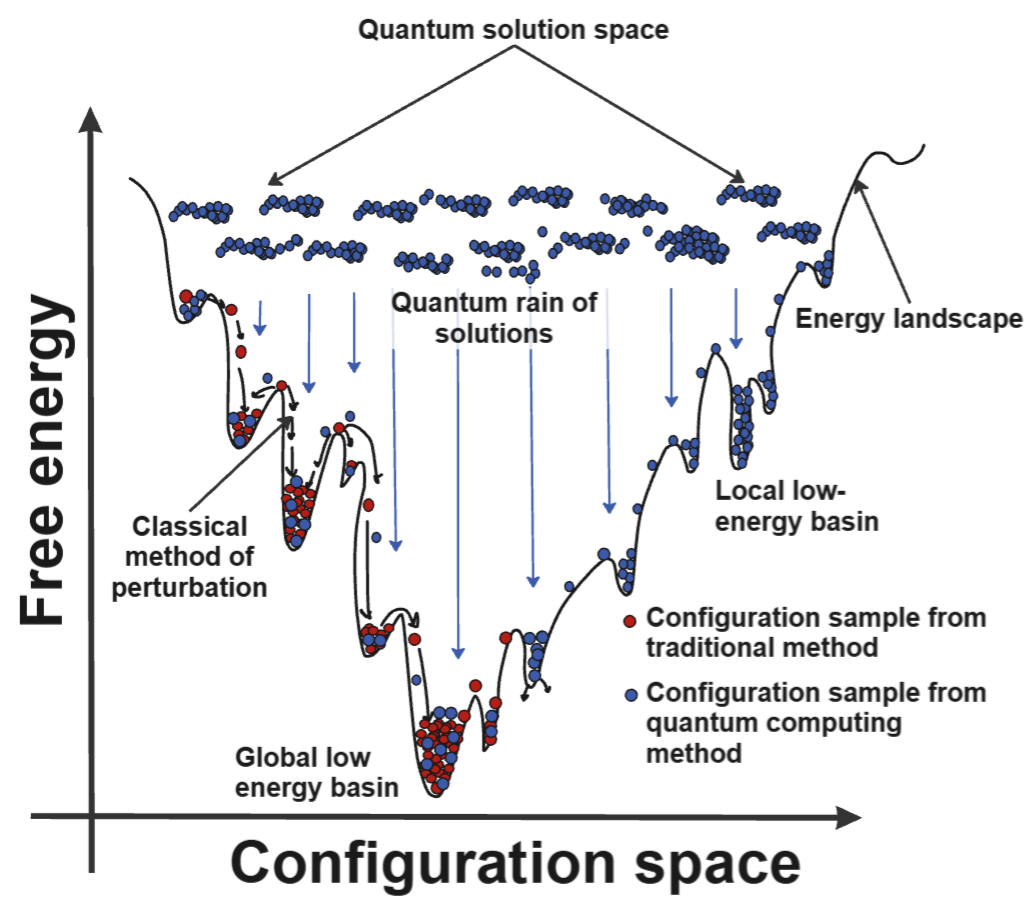}
    \caption{
    \textbf{Conceptual illustration of quantum-circuit ensemble sampling.}
    Superposition represents possible binary residue states, while structure-derived entangling operations introduce correlations between interacting residues. Repeated measurements generate residue-level microstates that are mapped to an energy distribution. These states represent thermodynamic residue configurations rather than alternative three-dimensional structures or dynamical trajectories.
    }
    \label{fig:conceptual_sampling_landscape}
\end{figure}

This framework shows that a residue-level quantum circuit can map a static protein structure to a coarse-grained thermodynamic ensemble. The sampled Gibbs-energy distribution captures intrinsic heterogeneity in the protein energy landscape and can be interpreted as reflecting folding-funnel organization, as illustrated in Fig.~\ref{fig:trp_energy_funnel}. Unlike lattice-based conformational searches or molecular dynamics simulations, which build ensembles through discrete structural enumeration or time evolution on a potential-energy surface, our approach generates each sample directly from a structure-conditioned probability landscape. Repeated circuit executions therefore produce a quantum “rain” of thermodynamic microstates, yielding a discrete population distribution rather than a continuous trajectory, as illustrated in Fig.~\ref{fig:conceptual_sampling_landscape}. Consequently, frequently sampled microstates represent more highly populated regions of the encoded energy landscape, which is usually found in convexity of the energy landscape. Beyond the energy landscape, the ensemble provides residue-level insight into the underlying local energetics. Here, information gain highlights residues whose state changes substantially reshape the distribution. This can also help to establish transitions between metastable states if we consider the ensembles of more protein structures (NMR conformations) and stitch them together. Further, pairwise and higher-order coupling maps (Dimer/Trimer level couplings) help identify cooperative residue groups that may potentially correspond to structural anchors, packing motifs, hinge regions, binding-site networks, and other distal regulations. Strong couplings between distal motifs and active-site residues may indicate potential allosteric sites. Subsequently, comparative analysis across apo, ligand-bound, wild-type, and mutant structures can then reveal changes in local and global stability, identify sensitive residues, and expose altered long-range couplings. Overall, the framework establishes a foundation for quantum-circuit-based protein ensemble modelling and provides a novel framework for extracting biophysical interpretations from structure-conditioned thermodynamic distributions

From a computational standpoint, a further motivation for the framework is its scaling behaviour relative to established simulation paradigms such as all-atom molecular dynamics and classical Monte Carlo sampling. Molecular dynamics simulations, whether all-atom or coarse-grained, propagate the system along a continuous trajectory governed by equations of motion, with a cost per time step scaling as $\mathcal{O}(n_{\mathrm{atom}}^2)$. The number of time steps required to sample thermodynamically relevant transitions is, however, determined not by this per-step cost but by the height of the free-energy barriers separating conformational basins: for a barrier of height $\Delta G^\ddagger$, the mean first-passage time scales exponentially as $\sim\exp(\Delta G^\ddagger/k_BT)$~\cite{Peters2024ProteinLandscape}. In practice, folding--unfolding transitions of even small proteins such as Trp-cage require nanosecond-scale trajectories with femtosecond integration steps, corresponding to $n_{\mathrm{step}}\sim10^6$--$10^8$. Enhanced sampling techniques such as replica-exchange molecular dynamics or meta-dynamics reduce effective barrier heights, and machine-learning-guided adaptive schemes have further improved sampling efficiency in recent years~\cite{Kleiman2023Adaptive}, but these introduce additional replicas, bias potentials, or learned collective variables rather than removing the underlying scaling constraint~\cite{Ghamari2022QuantumTransitions}. Successive frames along a trajectory are moreover highly autocorrelated, so a trajectory of length $T_{\mathrm{sim}}$ yields only $T_{\mathrm{sim}}/\tau_{\mathrm{corr}}$ statistically independent samples, where the autocorrelation time $\tau_{\mathrm{corr}}$ is itself proportional to the barrier-crossing time.An alternative is to sample binary residue configurations $\mathbf{s}\in\{0,1\}^N$ directly using Markov chain Monte Carlo (MCMC), with the residue Hamiltonian defining acceptance probabilities. The cost per step is modest ($\mathcal{O}(N)$ for a single-spin-flip Metropolis update~\cite{Metropolis1953}) and memory requirements are polynomial, but the effective sampling efficiency is governed by the mixing time $\tau_{\mathrm{mix}}$, the number of steps required for the chain to converge to its stationary distribution from an arbitrary initial state. On the rugged energy landscapes characteristic of protein thermodynamics, where multiple local minima are separated by high barriers, $\tau_{\mathrm{mix}}$ can grow exponentially with system size or barrier height~\cite{Dill2012ProteinFolding,Swendsen1986}. The energy distribution obtained for Trp-cage (Fig.~\ref{fig:trp_energy_funnel}) exhibits precisely this structure: multiple discrete low-energy peaks separated by energetic gaps indicate that an MCMC sampler would become trapped in individual sub-basins and require long excursions to visit competing minima. Parallel tempering and simulated tempering can accelerate inter-basin mixing~\cite{Swendsen1986,Marinari1992}, but require maintaining multiple coupled chains at different temperatures, increasing total cost by a factor proportional to the number of replicas. It is further tempting to ask whether a classical probabilistic graphical model could reproduce the same correlated residue-state distribution without recourse to quantum hardware. Bayesian networks are defined over a directed acyclic graph (DAG)~\cite{Pearl1988}, and molecular contact networks do not naturally admit such a structure: hydrogen bonds, salt bridges, and packing contacts are symmetric, mutual interactions between residues rather than hierarchical causal dependencies, so encoding them as a DAG would require an arbitrary choice of edge direction and topological ordering with no physical justification. Undirected models such as Markov random fields remove the acyclicity requirement~\cite{GemanGeman1984}, but shift the cost elsewhere: lacking a tractable joint normalising constant, they must be sampled iteratively under an explicit noise or temperature schedule, which returns the problem to exactly the mixing-time limitation described above. Undirected models of this kind have nonetheless been applied successfully in protein modelling, most notably in Potts-model direct-coupling analysis of residue co-evolution~\cite{Morcos2011}. The point is therefore not that classical graphical models are infeasible, but that they incur the equilibration overhead that the present construction is designed to avoid. By contrast, the framework presented here prepares a correlated residue-state distribution directly through a structure-informed circuit and samples from it by repeated measurement, avoiding explicit time integration, barrier-crossing trajectories, and Markov-chain equilibration entirely; each shot is an independent draw, so no autocorrelation or convergence diagnostic is required. The computational cost of our method is instead determined by the number of residues, encoded interactions, circuit layers, and measurement shots, with the state represented using $\mathcal{O}(N \times M )$ qubits on a fault tolerant quantum hardware, where $N$ is the number of residues and $M$ is the number of shots. Although exact classical simulation of a general $N$-qubit circuit requires storing $\mathcal{O}(2^N)$ amplitudes, this bound is far from tight for structured circuits: tensor-network methods matched to the interaction geometry can simulate shallow, sparsely connected circuits at substantially lower cost~\cite{Tindall2024Eagle,Begusic2024Converged,Rudolph2025TensorSampling}, and at the system sizes considered here classical simulation remains entirely tractable. Any quantum advantage would therefore be expected only for larger systems whose contact topology is sufficiently connected that the resulting output distribution resists efficient classical approximation. Overall, the practical value of the framework at present therefore lies in providing a compact, physically interpretable generative model for residue-level solvation ensemble on a quantum computer.

\subsection{Potential Enhancements}
\label{sec:disc-enhancements}

The circuit considered in this research only considered a single structure for the ansatz, while the NMR structures provide more variations in the protein structures, it should be incorporated to encode the ansatz on the circuit. At this point, we can think of a variational algorithm to generate the samples using multiple such structures, and shall be explored in the subsequent work. Further, we believe that the statistics generated from the circuit can be used in many ways and contain a lot of information such as fragment based couplings, that can further reveal interesting insights into protein-protein interactions. More importantly, the perturbations/solvation irritation caused by the cofactors or the environment would be an interesting application to explore. At this point, this method opens up so many of avenues to explore from applications as well as quantum circuit based modelling. However, as the quantum circuit for this study is simulated on the GPU, the next step going forward would be to produce the results on a real quantum computers. We didn't execute on a real quantum circuit at this stage because it comes with following challenges. The execution on a real quantum hardware would introduce gate errors, decoherence, and readout bias, all of which can reshape the sampled distribution and alter inferred dependencies. We rely on pennylane tool for hardware based error corrections, but given the dynamic research exploration in quantum systems, the error correction is also rapidly progressing. Moreover, it may also be required to think of new calibration methods for such samplings on a real Quantum Computer \cite{bergholm2022}. Another point to consider is the high number of samples we needed to show the pattern, however it is to be noted that it depends on the user confidence on the quantum-circuit's generated distribution, also a real quantum computer can do a much better job for sampling a high quality noise and may require lesser samples to reveal the true distribution. Further, we also note that the residue representation is binary and two-level variable captures an effective stabilised-versus-excited distinction but cannot resolve multiple rotameric states, continuous backbone dihedral angles, or explicit secondary-structure order parameters. Extending the representation to multi-level encodings or multiple qubits per residue would further enhance the results and deeper insights into dynamic nature of proteins. Furthermore, we note that quantitative scaling advantage of quantum hardware execution assumes fault-tolerant quantum processors with sufficient gate fidelity and qubit connectivity to implement the required interaction graphs without prohibitive compilation overhead. Current noisy intermediate-scale quantum (NISQ) devices impose additional constraints such as limited qubit counts, restricted connectivity topologies, and finite coherence times, that would restrict the practical protein size and circuit depth achievable at useful precision. The transition from the noiseless simulations reported here to hardware execution will therefore require systematic characterisation of noise-induced biases in ensemble observables and development of circuit-specific error mitigation protocols. Lastly, we note that the results generated by the quantum circuit require careful validation, with thoughtful selection of validation data and methods. Going forward, strategies such as fragmentation may prove useful, and the model could be further improved by drawing inspiration from Fragment Molecular Orbital theory \cite{Fedorov2017}.

\section*{Code Availability}
\label{code_avail}
The source code, configuration files, and analysis scripts used in this study are publicly available at \url{https://github.com/pra-ashok/QProtSim}. QProtSim is also available as a pypi package: \texttt{pip install qprotsim}.

\section*{Acknowledgement}

The authors, PP and BB thank the University of Suffolk and Evotec Pharma, UK for supporting this research. Pratik Patil received his doctoral funding through the University of Suffolk-Evotec partnership, with the Evotec Pharma sponsored PhD scholarship. BB was formally a full-time employee of Evotec Pharma.

\section*{AI usage}

We used large language models (GPT-5) to assist in improving the grammar of the manuscript, as well as facilitate code documentation.

\bibliographystyle{IEEEtran}
\bibliography{main}

\clearpage        
\onecolumn        
\appendix
\section{Appendix: Estimators, statistical validation, and reproducibility}
\label{app:estimators}

\setcounter{figure}{0}
\counterwithin{figure}{section}

\setcounter{table}{0}
\counterwithin{table}{section}

\setcounter{equation}{0}
\counterwithin{equation}{subsection}

\subsection{Binary-state convention and measurement mapping}
\label{app:state_mapping}
Our circuit measurements return per-residue outcomes in the Pauli-$Z$ basis, $Z_r \in \{+1,-1\}$ for residue $r$. We map these outcomes to a binary excitation indicator $s_r \in \{0,1\}$ via

\begin{align}
s_r &= \mathbb{I}\!\left[ Z_r = z_{\mathrm{exc}} \right], \qquad
z_{\mathrm{exc}} \in \{+1,-1\}.
\label{eq:state_map}
\end{align}

Here $z_{\mathrm{exc}}$ denotes the measurement eigenvalue identified with the ``excited'' state in our encoding (fixed throughout all experiments). For a set of $N$ circuit shots, we denote the sampled excitation matrix as $\mathbf{S}\in\{0,1\}^{N\times n}$, where $n$ is the number of residues/qubits.

\subsection{Probability-to-Angle Encoding}
\label{app:encoding_angle}

A qubit stores measurement probabilities in the squared magnitudes of its amplitudes. For

\begin{align}
    \ket{\psi}
    &=
    a\ket{0}
    +
    b\ket{1},
    &
    |a|^2+|b|^2
    &=
    1,
\end{align}
the Born rule gives
\begin{align}
    P(0)
    &=
    |a|^2,
    &
    P(1)
    &=
    |b|^2.
\end{align}
Thus, a desired probability $p$ for observing $\ket{1}$ is represented by
\begin{align}
    \ket{\psi(p)}
    &=
    \sqrt{1-p}\ket{0}
    +
    \sqrt{p}\ket{1}.
    \label{eq:probability_state}
\end{align}

This state can be prepared from $\ket{0}$ using a rotation about the $y$-axis of the Bloch sphere. With the standard convention
\cite{NielsenChuang2010,IBMQuantumRYGate},
\begin{align}
    R_Y(\theta)
    &=
    \begin{pmatrix}
        \cos(\theta/2) & -\sin(\theta/2) \\
        \sin(\theta/2) &  \cos(\theta/2)
    \end{pmatrix},
    \label{eq:ry_matrix}
\end{align}
and therefore
\begin{align}
    R_Y(\theta)\ket{0}
    &=
    \cos\!\left(\frac{\theta}{2}\right)\ket{0}
    +
    \sin\!\left(\frac{\theta}{2}\right)\ket{1}.
    \label{eq:ry_action}
\end{align}
The probability of measuring the excited basis state is
\begin{align}
    P(1)
    &=
    \sin^2\!\left(\frac{\theta}{2}\right).
    \label{eq:ry_probability}
\end{align}
Setting $P(1)=p$ gives the inverse probability-to-angle map
\begin{align}
    p
    &=
    \sin^2\!\left(\frac{\theta}{2}\right),
    &
    \theta
    &=
    2\arcsin\sqrt{p}.
    \label{eq:probability_angle_inverse}
\end{align}
The factor of two follows directly from the half-angle dependence of $R_Y(\theta)$.

In the present model, each residue $r$ is assigned an excitation probability $p_r$. The corresponding qubit is initialised with

\begin{align}
    \theta_r
    &=
    2\arcsin\sqrt{p_r},
    \label{eq:residue_probability_angle}
\end{align}
which prepares
\begin{align}
    \ket{\psi_r}
    &=
    \sqrt{1-p_r}\ket{0}
    +
    \sqrt{p_r}\ket{1}.
    \label{eq:residue_initial_state}
\end{align}
With $\ket{0}$ denoting the residue ground state and $\ket{1}$ denoting the excited state,
\begin{align}
    P_r(\mathrm{ground})
    &=
    1-p_r,
    &
    P_r(\mathrm{excited})
    &=
    p_r.
    \label{eq:residue_measurement_probabilities}
\end{align}

The same mapping applies to a controlled-$R_Y$ operation. A controlled rotation applies $R_Y(\phi)$ to the target only when the control qubit is active:

\begin{align}
    CR_Y(\phi)
    &=
    \ket{0}\!\bra{0}_c \otimes I_t
    +
    \ket{1}\!\bra{1}_c \otimes R_Y(\phi)_t.
    \label{eq:cry_operator}
\end{align}
For an active control and a target initially in $\ket{0}$,
\begin{align}
    CR_Y(\phi)\ket{10}
    &=
    \cos\!\left(\frac{\phi}{2}\right)\ket{10}
    +
    \sin\!\left(\frac{\phi}{2}\right)\ket{11}.
    \label{eq:cry_action}
\end{align}
Hence the conditional target-excitation probability is
\begin{align}
    P(t=1 \mid c=1)
    &=
    \sin^2\!\left(\frac{\phi}{2}\right).
    \label{eq:cry_conditional_probability}
\end{align}
A conditional transition probability $q$ is therefore encoded as
\begin{align}
    q
    &=
    \sin^2\!\left(\frac{\phi}{2}\right),
    &
    \phi
    &=
    2\arcsin\sqrt{q}.
    \label{eq:cry_probability_angle}
\end{align}

Therefore, $R_Y$ encodes a single-residue excitation probability, while controlled-$R_Y$ encodes a conditional transition probability between two residue qubits.

\subsection{Gate ordering sensitivity}
\label{app:gate_ordering}

Interaction gates are applied sequentially, and their order may affect the final quantum state and the resulting measurement distribution. This dependence arises when two gate operations do not commute. Consider two interaction gates, \(U_a\) and \(U_b\), within the entanglement block. They commute when applying them in either order gives the same operation:

\begin{align}
    U_a U_b = U_b U_a,
\end{align}

or, equivalently,

\begin{align}
    [U_a,U_b]
    =
    U_aU_b-U_bU_a
    =
    0.
\end{align}

When the above commute condition is satisfied, exchanging the positions of the gates does not alter the circuit output. Gates acting on disjoint qubit pairs always commute. Gates sharing a qubit may also commute if their operations are compatible. In contrast, overlapping gates that apply different non-commuting operations to the shared qubit may not commute, i.e. $U_aU_b \neq U_bU_a$. Hence, their order can then change the final state and, consequently, the probability of measuring a residue-state configuration \(\mathbf{Z}\):

\begin{align}
    P(\mathbf{Z})
    =
    \left|
    \braket{\mathbf{Z}|\psi_{\mathrm{final}}}
    \right|^2.
\end{align}

Gate ordering is therefore part of the interaction-circuit specification. Its influence can be reduced by grouping non-overlapping interactions into parallel layers, using a fixed ordering rule across all systems, and dividing large rotations across repeated layers. A symmetric forward--reverse ordering may also reduce directional bias introduced by a single sequential ordering. Moreover, ordering robustness can be assessed by repeating the calculation with reversed or alternative valid layer orders and comparing the resulting residue occupancies, energy distributions, and coupling matrices. On quantum hardware, circuit depth and additional SWAP operations should also be minimised so that ordering sensitivity is not amplified by device noise.

\subsection{Residue transfer-energy reference tables}
\label{app:transfer_energies}

\begin{table}[htbp]
    \centering
    \scriptsize
    \setlength{\tabcolsep}{4pt}
    \renewcommand{\arraystretch}{1.12}
    \caption{\textbf{Residue transfer-energy references implemented in
    \texttt{circuit.py}.} \(E_1=0\) for both references, so
    \(\Delta G=-E_0\).}
    \label{tab:transfer_energy_refs}
    \begin{tabular}{llrrrr}
        \hline
        \textbf{Residue} & \textbf{Name} &
        \(\boldsymbol{E_0^{\mathrm{bulk}}}\) &
        \(\boldsymbol{\Delta G^{\mathrm{bulk}}}\) &
        \(\boldsymbol{E_0^{\mathrm{alkyl}}}\) &
        \(\boldsymbol{\Delta G^{\mathrm{alkyl}}}\) \\
        \hline
        A & ALA & -1.76 & 1.76 & -2.67 & 2.67 \\
        R & ARG & -5.72 & 5.72 & -1.75 & 1.75 \\
        N & ASN & -3.40 & 3.40 & -1.25 & 1.25 \\
        D & ASP & -4.36 & 4.36 & -1.03 & 1.03 \\
        C & CYS & -8.73 & 8.73 & -4.05 & 4.05 \\
        E & GLU & -3.63 & 3.63 & -3.00 & 3.00 \\
        Q & GLN & -1.25 & 1.25 & -1.95 & 1.95 \\
        G & GLY & -0.10 & 0.10 & -0.10 & 0.10 \\
        H & HIS & -0.74 & 0.74 & -4.22 & 4.22 \\
        I & ILE & -10.20 & 10.20 & -11.40 & 11.40 \\
        L & LEU & -9.64 & 9.64 & -8.95 & 8.95 \\
        K & LYS & -5.61 & 5.61 & -3.00 & 3.00 \\
        M & MET & -6.97 & 6.97 & -8.23 & 8.23 \\
        F & PHE & -10.10 & 10.10 & -12.30 & 12.30 \\
        P & PRO & -4.08 & 4.08 & -7.15 & 7.15 \\
        S & SER & -0.23 & 0.23 & -0.42 & 0.42 \\
        T & THR & -1.47 & 1.47 & -2.50 & 2.50 \\
        W & TRP & -12.80 & 12.80 & -11.30 & 11.30 \\
        Y & TYR & -5.44 & 5.44 & -8.70 & 8.70 \\
        V & VAL & -6.91 & 6.91 & -7.80 & 7.80 \\
        \hline
    \end{tabular}
\end{table}

The circuit defines two residue transfer-energy references. The final bulk-octanol runs use the Fauchere--Pliska bulk-octanol table by default, whereas the alkyl-chain table is available through \texttt{energies\_ref("alkyl")} as the reversed-phase liquid-chromatography stationary-phase reference. In both code tables, the excited reference energy is fixed at \(E_1=0\) for every residue. The transfer-energy gap used by the directed-transfer schedule is therefore

\begin{align}
    \Delta G = E_1 - E_0 = -E_0.
\end{align}

Table~\ref{tab:transfer_energy_refs} lists the implemented \(E_0\) values and
the corresponding positive gaps for both references.


\subsection{Mutual information estimator for residue pairs}
\label{app:mi}
For a residue pair $(i,j)$ and binary states $a,b\in\{0,1\}$, we estimate joint and marginal probabilities from samples:
\begin{align}
    \hat{p}_{ij}(a,b) &= \frac{1}{N}\sum_{k=1}^{N}\mathbb{I}\!\left[s_i^{(k)}=a,\, s_j^{(k)}=b\right], \\
    \hat{p}_{i}(a) &= \sum_{b\in\{0,1\}} \hat{p}_{ij}(a,b), \qquad
    \hat{p}_{j}(b) = \sum_{a\in\{0,1\}} \hat{p}_{ij}(a,b).
\end{align}
The (Shannon) mutual information is then
\begin{align}
    \mathrm{MI}(i,j)
    &= \sum_{a\in\{0,1\}}\sum_{b\in\{0,1\}}
    \hat{p}_{ij}(a,b)\,
    \log \!\left(\frac{\hat{p}_{ij}(a,b)}{\hat{p}_{i}(a)\,\hat{p}_{j}(b)}\right),
    \label{eq:mi_estimator}
\end{align}
with $\log(\cdot)$ denoting the natural logarithm (units: nats). If $\log_2$ is used instead, MI is reported in bits. In practice, terms with $\hat{p}_{ij}(a,b)=0$ are omitted from the sum.

\subsection{\texorpdfstring{$K_{ij}$}{Kij} Co-efficients}
\label{app:kij}

For binary excitation variables, we quantify linear association via the Pearson correlation on \(\{0,1\}\) indicators, equivalently the \(\phi\) coefficient. Let \(\rho_{ij}\) denote this correlation computed from samples. To avoid interpreting spurious correlations arising from finite-shot noise, we gate pairwise correlation analyses by testing whether dependence is statistically significant under an independence null. The correlation gate is therefore applied to sampled statistical dependence, whereas the structure-derived \(K_{ij}\) matrix is used as the structural prior that defines which residue pairs can exchange excitation in the quantum circuit and as structural context for interpreting gated correlations.

For each residue pair \((i,j)\), \(K_{ij}\) is constructed from bounded structural descriptors \(\kappa^{x}_{ij}\in[0,1]\). These descriptors encode peptide connectivity, disulfide bonding, hydrogen bonding, salt-bridge contacts, and short-range packing. They are first combined into an unnormalised structural stiffness score,
\begin{align}
    S_{ij}
    =
    w_{\mathrm{BB}}\kappa^{\mathrm{BB}}_{ij}
    +
    w_{\mathrm{SS}}\kappa^{\mathrm{SS}}_{ij}
    +
    w_{\mathrm{HB}}\kappa^{\mathrm{HB}}_{ij}
    +
    w_{\mathrm{salt}}\kappa^{\mathrm{salt}}_{ij}
    +
    w_{\mathrm{vdW}}\kappa^{\mathrm{vdW}}_{ij}.
\end{align}
The weights used for the final runs were
\begin{align}
    w_{\mathrm{BB}}=1.00,\quad
    w_{\mathrm{SS}}=1.00,\quad
    w_{\mathrm{HB}}=0.50,\quad
    w_{\mathrm{salt}}=0.40,\quad
    w_{\mathrm{vdW}}=0.05.
\end{align}
These weights are relative coupling priors, not absolute bond, based on their relative strengths. Peptide and disulfide terms receive the largest weights because they impose direct covalent constraints. Hydrogen bonds and salt bridges receive intermediate weights because they are stabilising but geometry- and environment-dependent. The van der Waals weight is deliberately small because packing contacts are numerous and individually weak.

The backbone term preserves covalent chain connectivity. Adjacent residues receive unit coupling when the peptide C--N distance is compatible with a peptide bond,
\begin{align}
    \kappa^{\mathrm{BB}}_{ij}
    =
    \begin{cases}
        1, & |i-j|=1\ \mathrm{and}\ r_{\mathrm{C}_i,\mathrm{N}_j}
        \leq 1.8~\text{\AA},\\
        0, & \mathrm{otherwise}.
    \end{cases}
\end{align}
For incomplete structures lacking the relevant C/N atoms, sequence adjacency is used as the fallback. The disulfide term is a long-range covalent crosslink term,
\begin{align}
    \kappa^{\mathrm{SS}}_{ij}
    =
    \begin{cases}
        1, & r_{\mathrm{SG}_i,\mathrm{SG}_j} \leq 2.3~\text{\AA},\\
        0, & \mathrm{otherwise},
    \end{cases}
\end{align}
for cysteine-like residue pairs. Hydrogen-bond coupling is evaluated using a heavy-atom approximation over polar atoms and a \(3.5~\text{\AA}\) cutoff,
\begin{align}
    \kappa^{\mathrm{HB}}_{ij}
    =
    1 -
    \exp\left[
    -
    \sum_{a\in i,\,b\in j}
    \left(\frac{3.5}{r_{ab}}\right)^{2}
    \right],
    \qquad r_{ab}\leq 3.5~\text{\AA}.
\end{align}
Salt-bridge coupling uses oppositely charged side-chain atoms within
\(4.0~\text{\AA}\),
\begin{align}
    \kappa^{\mathrm{salt}}_{ij}
    =
    1 -
    \exp\left[
    -
    \sum_{a\in i,\,b\in j}
    \left(\frac{4.0}{r_{ab}}\right)^{2}
    \right],
    \qquad r_{ab}\leq 4.0~\text{\AA}.
\end{align}
Short-range packing is represented by a softened heavy-atom contact score,
\begin{align}
    \kappa^{\mathrm{vdW}}_{ij}
    =
    1 -
    \exp\left[
    -
    \frac{1}{8}
    \sum_{a\in i,\,b\in j}
    \left(\frac{3.5}{r_{ab}}\right)^{2}
    \right],
    \qquad r_{ab}\leq 4.0~\text{\AA}.
\end{align}

where $r_{ab}$ represents the distance between two atoms.

The structural stiffness \(S_{ij}\) is then modulated by local frequency matching and residue-mass compatibility. When explicit frequencies are not provided, a relative local frequency is estimated from the local stiffness and residue mass \(M_i\),

\begin{align}
    \omega_i =
    \sqrt{
    \frac{\sum_j S_{ij}}{M_i}
    }.
\end{align}

Pairwise frequency matching is
\begin{align}
    \Phi_{ij}
    =
    \exp\left[
    -\frac{(\omega_i-\omega_j)^2}{2\sigma_\omega^2}
    \right],
\end{align}

where \(\sigma_\omega\) is estimated from the standard deviation of the relative frequencies when not supplied explicitly. The mass compatibility factor is

\begin{align}
    \Lambda_{ij}^{\mathrm{mass}}
    =
    \frac{2\sqrt{M_iM_j}}{M_i+M_j},
\end{align}
which equals one for equal masses and decreases smoothly as residue masses diverge. The unnormalised coupling is therefore

\begin{align}
    K_{ij}^{\mathrm{raw}}
    =
    S_{ij}\Phi_{ij}\Lambda_{ij}^{\mathrm{mass}}.
\end{align}
All positive pair couplings are finally normalised by the maximum observed raw coupling,

\begin{align}
    K_{ij}
    =
    \frac{K_{ij}^{\mathrm{raw}}}
    {\max_{a<b} K_{ab}^{\mathrm{raw}}},
    \qquad
    0 \leq K_{ij}\leq 1.
\end{align}

This normalisation preserves the rank order of structural interaction strengths while making the coupling scale comparable between 1L2Y and 9GDL.

In the circuit, \(K_{ij}\) enters the directed excitation-transfer schedule. For local excitation propensity \(p_i\), energy gap
\(\Delta G_i=E_{1,i}-E_{0,i}\), and thermodynamic scale \(\beta\), the local drive variable is

\begin{align}
    \mu_i=\beta\Delta G_i p_i.
    \label{eq:exc-pot}
\end{align}

The positive directed drive from residue \(i\) to residue \(j\) is

\begin{align}
    \nu_{i\rightarrow j}
    =
    \max\left[
    0,\,
    \frac{\mu_i-\mu_j}{\mu_i+\mu_j+\epsilon}
    \right],
\end{align}
with a small numerical stabiliser in the denominator. For each source residue, the local coupling denominator is

\begin{align}
    D_i=\sum_{k\in\mathcal{N}(i)}K_{ik}.
\end{align}

The corresponding implemented transfer probability is
\begin{align}
    \Delta P_{i\rightarrow j}
    =
    \frac{2K_{ij}\nu_{i\rightarrow j}}
    {D_i+D_j+\epsilon}
\end{align}
Thus \(K_{ij}\) controls how strongly structural contact can transmit excitation, while the sampled correlation and mutual-information analyses test whether statistically meaningful residue-state dependence is actually observed after measurement.

\subsection{Dimerisation and trimerisation motif analyses}
\label{app:motifs}

\begin{figure}[ht]
    \centering
    \includegraphics[width=0.78\textwidth]{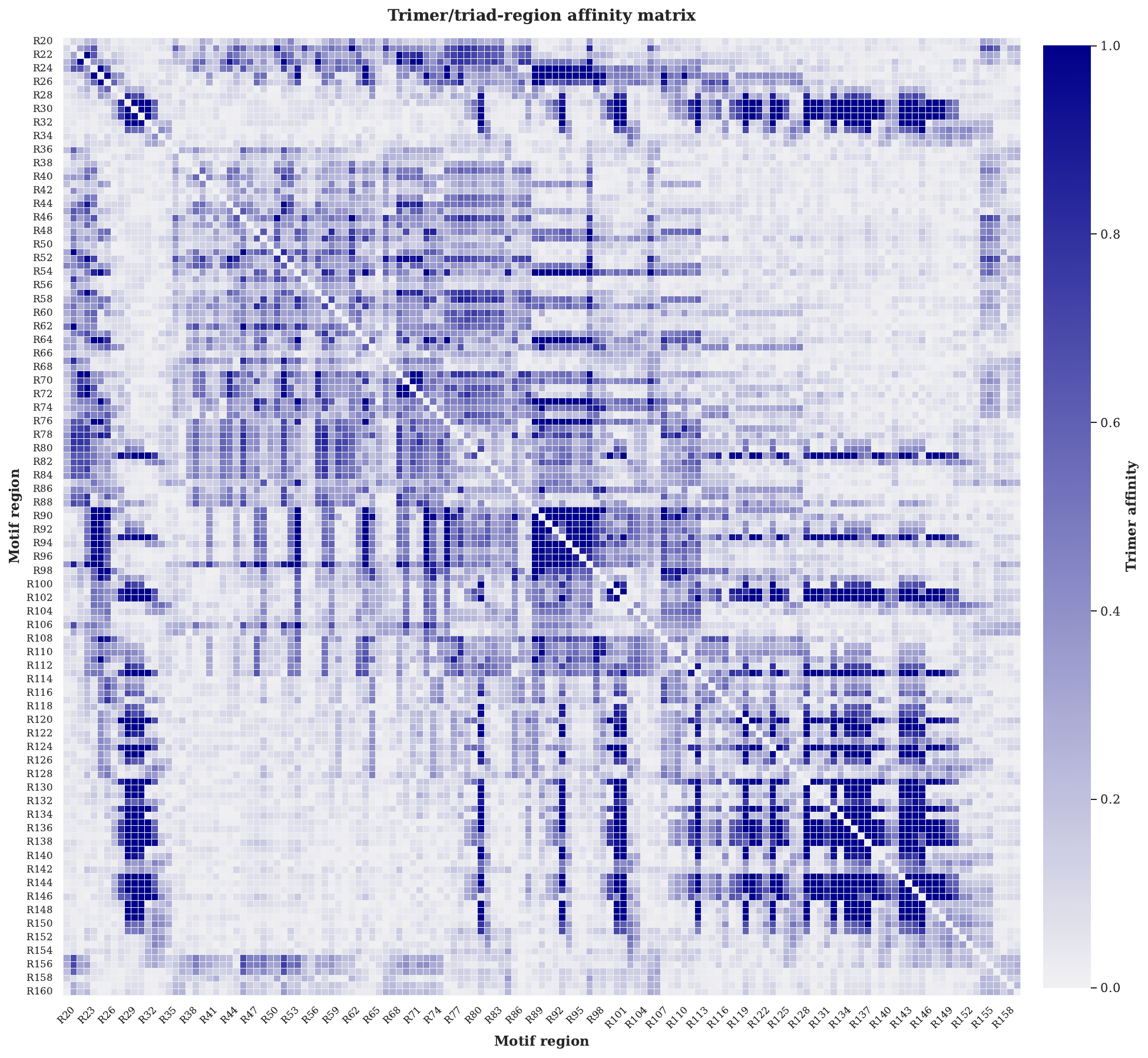}
    \caption{Trimer/triad-level allostery affinity matrix for 1L2Y. Each axis
    indexes consecutive peptide trimers and peptide/non-covalent triads,
    showing how residue-level directional excitation effects combine into
    three-residue motif communication maps.}
    \label{fig:app_trimer_affinity_1l2y}
\end{figure}

To analyse allosteric organisation above the single-residue level, residues were grouped into overlapping dimer and trimer/triad motif regions. These motifs are structural regions derived from the same residue-residue contact decomposition used to construct \(K_{ij}\). They should therefore be interpreted as local or semi-local residue groups over which measured directional excitation effects are aggregated, not as new quantum observables or separate simulations.

\begin{figure}[ht]
    \centering
    \includegraphics[width=0.78\textwidth]{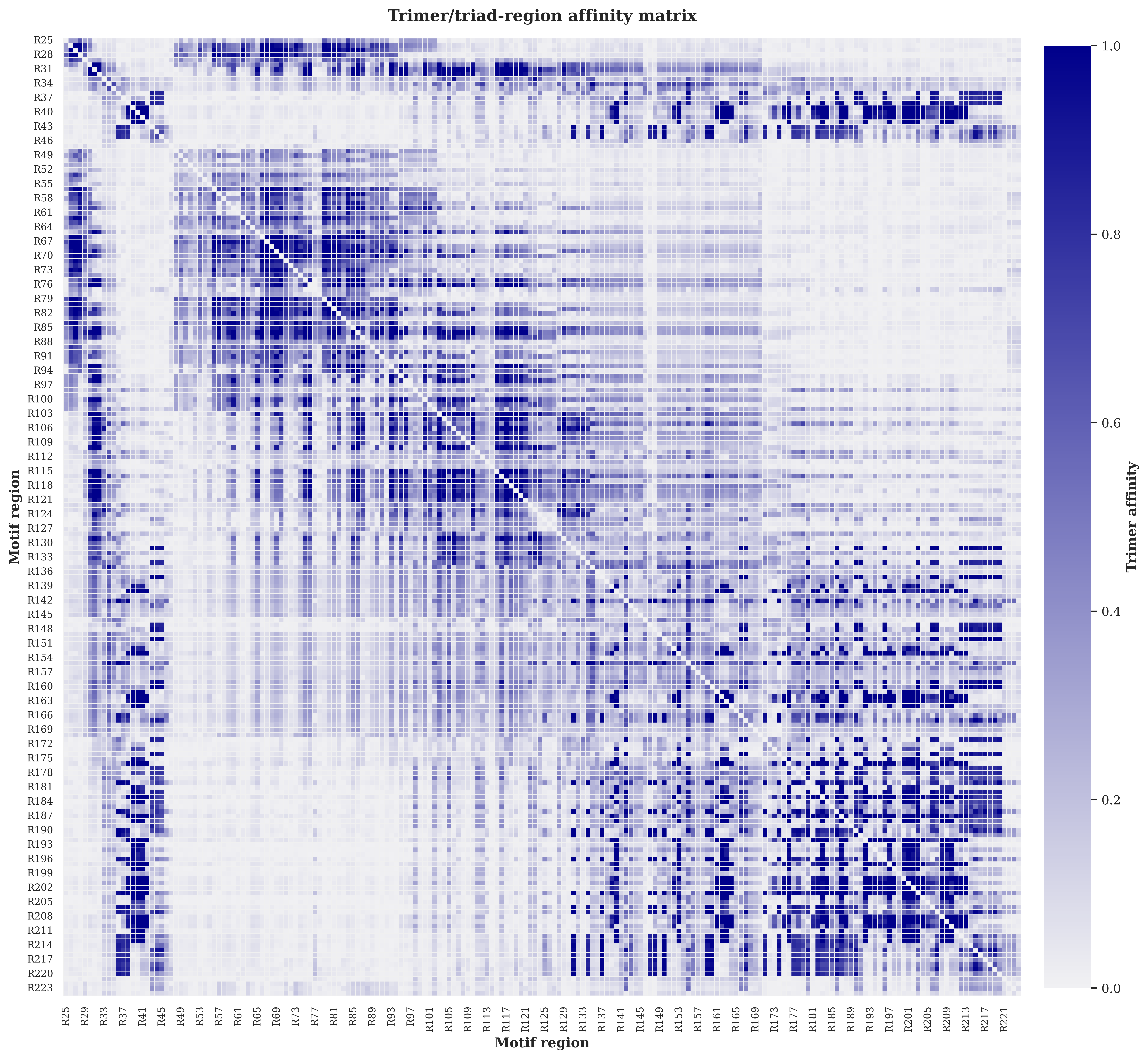}
    \caption{Trimer/triad-level allostery affinity matrix for 9GDL. Each axis
    indexes consecutive peptide trimers and peptide/non-covalent triads,
    showing how residue-level directional excitation effects combine into
    three-residue motif communication maps.}
    \label{fig:app_trimer_affinity_9gdl}
\end{figure}

Peptide dimer regions are residue pairs connected by the backbone term in the
\(K_{ij}\) model,
\begin{align}
    \mathcal{D}
    =
    \left\{
    \{i,j\}:
    |i-j|=1,\,
    \kappa^{\mathrm{BB}}_{ij}>0
    \right\}.
\end{align}

Thus each dimer represents a peptide-supported residue pair. Consecutive peptide trimers are three-residue windows in which both peptide edges are present,
\begin{align}
    \mathcal{T}_{\mathrm{pep}}
    =
    \left\{
    \{i,i+1,i+2\}:
    \kappa^{\mathrm{BB}}_{i,i+1}>0,\,
    \kappa^{\mathrm{BB}}_{i+1,i+2}>0
    \right\}.
\end{align}
Additional trimer/triad regions are allowed when three residues form a peptide-supported and non-covalently supported structural motif. Let

\begin{align}
    \eta_{ij}
    =
    \mathbf{1}
    \left(
    \kappa^{\mathrm{HB}}_{ij}
    +
    \kappa^{\mathrm{salt}}_{ij}
    +
    \kappa^{\mathrm{vdW}}_{ij}
    >0
    \right)
\end{align}

denote non-covalent support. A peptide/non-covalent triad is then a three-residue set with at least one peptide edge and at least one
non-covalent edge,

\begin{align}
    \mathcal{T}_{\mathrm{triad}}
    =
    \left\{
    \{i,j,k\}:
    \sum_{(a,b)\in E_{ijk}}
    \mathbf{1}\left(\kappa^{\mathrm{BB}}_{ab}>0\right)\geq 1,\,
    \sum_{(a,b)\in E_{ijk}}\eta_{ab}\geq 1
    \right\},
\end{align}
where \(E_{ijk}=\{(i,j),(i,k),(j,k)\}\). Disulfide contacts contribute to \(K_{ij}\) as covalent constraints, but they are not counted as
non-covalent support in this triad definition.

Because motifs overlap, a residue can contribute to multiple regions. The membership matrix \(W\) assigns residue \(i\) to motif region \(r\),

\begin{align}
    W_{ir}
    =
    \begin{cases}
        m_i^{-1}, & i\in r,\\
        0, & i\notin r,
    \end{cases}
    \qquad
    m_i=\sum_r\mathbf{1}(i\in r).
\end{align}
This gives the per-residue normalisation

\begin{align}
    \sum_r W_{ir}=1
\end{align}

for residues that belong to at least one motif. Residue-level directional effects are computed from the measured binary excitation ensemble. For target residue \(i\) and excited source residue \(j\),

\begin{align}
    D_{i,j}
    =
    P(R_i=1\mid R_j=1)
    \log
    \frac{P(R_i=1\mid R_j=1)+\epsilon}
    {P(R_i=1)+\epsilon}.
\end{align}

Source residues with fewer than 20 excited samples are excluded from this aggregation. Directional motif coupling from source motif \(q\) to target motif \(r\) is then the weighted mean of the residue-level directional effects,

\begin{align}
    C_{r,q}
    =
    \frac{
    \sum_{i,j} W_{ir}W_{jq}D_{i,j}
    }{
    \sum_{i,j} W_{ir}W_{jq}+\epsilon
    }.
\end{align}

The final reported motif affinity is symmetric and magnitude-based,

\begin{align}
    A_{r,q}
    =
    |C_{r,q}|+|C_{q,r}|.
\end{align}
Positive affinities are scaled by the 95th percentile and clipped to \([0,1]\). This robust scaling preserves relative strong motif couplings
without allowing one extreme value to dominate the colour scale.

Using this construction, 1L2Y produced 160 active motif regions: 19 peptide dimers, 18 peptide trimers, and 123 peptide/non-covalent triads. 9GDL produced 224 active motif regions: 24 peptide dimers, 23 peptide trimers, and 177 peptide/non-covalent triads. The larger 9GDL system therefore has a larger motif graph and a larger distal motif-coupling search space. Distal motif pairs were defined as non-overlapping motif pairs that are not immediate sequence neighbours; equivalently, the nearest residue-residue sequence separation between the two motifs is at least two residues.

The dimer-level affinity matrices isolate peptide-pair communication. In 1L2Y, the strongest distal dimer signals are concentrated around Gly, Pro, and Ser pairs and links to Gln/Trp motifs. In 9GDL, the dominant distal dimer links shift toward acidic N-terminal pairs and the Tyr/Ile/Trp hydrophobic-aromatic region. The trimer/triad matrices show that this difference persists at the three-residue motif level (Fig.~\ref{fig:app_trimer_affinity_9gdl}). 1L2Y links Gln/Trp/Pro motifs with Asp/Gly/Ser motifs, whereas 9GDL links Glu/Cys/Ile motifs to the
central Arg/Leu/Trp region.

\subsection{Interaction layers}
\label{sec:layers}

One complete interaction layer applies gate updates on all peptide edges followed by all non-covalent edges. Within each edge class, a fixed deterministic ordering was used for all proteins and repeated runs. The complete interaction layer was executed only once, however it can be repeated ($L$) times:
\begin{align}
    U_{\mathrm{int}}
    &=
    \prod_{\ell=1}^{L}
    \left[
    U_{\mathrm{nc}}^{(\ell)}
    U_{\mathrm{pep}}^{(\ell)}
    \right],
\end{align}
where ($U_{\mathrm{pep}}^{(\ell)}$) and ($U_{\mathrm{nc}}^{(\ell)}$) denote the ordered products of peptide and non-covalent controlled rotations in layer ($\ell$), respectively.

Increasing ($L$) permits correlations to propagate over multi-edge pathways in the residue graph. Circuit depth is therefore treated as a model hyperparameter controlling interaction propagation. It is not assigned a direct physical timescale unless separately calibrated against dynamical data.

The final circuit state is given by
\begin{align}
    \ket{\psi_{\mathrm{final}}}
    &=
    U_{\mathrm{int}}
    \ket{\psi_{\mathrm{0}}}.
\end{align}

The layers may be repeated to increase the redistribution depth and allow multi-hop propagation of excitation across the residue network.

\subsection{Thermodynamic Consistency of the Circuit-Sampled Ensemble}
\label{app}

The central thesis is whether the circuit produces an arbitrary measurement distribution or a thermodynamically meaningful ensemble associated with the input protein structure. We establish the latter under three explicit modelling assumptions. First, the selected PDB structure is treated as an equilibrium representative of a folded conformational basin at temperature ($T$). Its atomic coordinates therefore define the structure-conditioned solvent exposure and residue-interaction network for that basin. Second, the residue excitation gaps represent effective free-energy costs for moving residues from their stabilised states to their excited solvation states. Third, the directed interaction block transfers excitation propensity from a residue with a larger local excitation potential to a structurally connected residue with a smaller excitation potential. Under these assumptions, the initial circuit encodes an independent Gibbs distribution, while the controlled interaction gates introduce correlations along a free-energy-descending direction.

\subsubsection{Exact Gibbs encoding before residue interactions}

Let
\begin{align}
    \mathbf{z}
    =
    (z_1,z_2,\ldots,z_N),
    \qquad
    z_r\in{0,1},
\end{align}
denote a residue-state configuration, where ($z_r=0$) represents the stabilised state and ($z_r=1$) represents the excited state of residue ($r$). The effective two-state free-energy score is

\begin{align}
    g_r(z_r)
    =
    (1-z_r)E_{0,r}^{\mathrm{eff}}
    +
    z_rE_{1,r}^{\mathrm{eff}},
\end{align}
with excitation gap

\begin{align}
    \Delta G_r^{\mathrm{eff}}
    =
    E_{1,r}^{\mathrm{eff}}
    -
    E_{0,r}^{\mathrm{eff}}.
\end{align}

At inverse temperature ($\beta=(k_{\mathrm B}T)^{-1}$), the equilibrium excitation probability is

\begin{align}
    p_r
    =
    \frac{
    \exp[-\beta\Delta G_r^{\mathrm{eff}}]
    }{
    1+\exp[-\beta\Delta G_r^{\mathrm{eff}}]
    }.
\end{align}
An ($R_Y$) rotation with
\begin{align}
    \theta_r
    =
    2\arcsin\sqrt{p_r}
\end{align}
prepares
\begin{align}
    \ket{\psi_r}
    =
    \sqrt{1-p_r}\ket{0}
    +
    \sqrt{p_r}\ket{1}.
\end{align}

Consequently, the initial ($N$)-qubit state is

\begin{align}
    \ket{\psi_0}
    =
    \bigotimes_{r=1}^{N}\ket{\psi_r},
\end{align}

and its measurement distribution is

\begin{align}
    P_0(\mathbf{z})
    =
    \prod_{r=1}^{N}
    (1-p_r)^{1-z_r}p_r^{z_r}
    \nonumber\
    =
    \frac{
    \exp[-\beta G_0(\mathbf{z})]
    }{
    Z_0
    },
\end{align}

where

\begin{align}
    G_0(\mathbf{z})
    =
    \sum_{r=1}^{N}g_r(z_r),
    \qquad
    Z_0
    =
    \sum_{\mathbf{z}}
    \exp[-\beta G_0(\mathbf{z})].
\end{align}
Thus, the single-qubit preparation exactly encodes the Gibbs distribution of the independent-residue model.

\subsubsection{Directed excitation transfer defines the optimisation direction}

The interaction graph determines the routes along which excitation may propagate, while the local free-energy gradient determines the direction of propagation. For each interacting residue pair $(i,j)$, the equilibrium PDB structure defines a non-negative coupling strength $K_{ij}\geq 0$. This coupling represents a structurally allowed pathway, such as a peptide bond, disulfide bond, hydrogen bond, salt bridge, or packing contact. A large $K_{ij}$ means that residues $i$ and $j$ are strongly connected, but it does not by itself define the direction of transfer.

Consider the probability vector $\mathbf{p}^{\prime}$ resulting from an accepted transfer, obtained by perturbing the entanglement structure through the redistribution of a probability by $\delta_{i\rightarrow j}$ from residue $i$ to residue $j$:

\begin{align}
    p_i^{\prime}
    =
    p_i
    -
    \delta_{i\rightarrow j},
    \qquad
    p_j^{\prime}
    =
    p_j
    +
    \delta_{i\rightarrow j}.
    \label{eq:probability-transfer}
\end{align}

To relate this update to the associated free-energy change, we introduce an intermediate path $\mathbf{p}(s)$ connecting the initial and final probability vectors:

\begin{align}
    p_i(s)
    =
    p_i
    -
    s,
    \qquad
    p_j(s)
    =
    p_j
    +
    s,
    \qquad
    0
    \leq
    s
    \leq
    \delta_{i\rightarrow j},
\end{align}
with all remaining probabilities held fixed.

Furthermore, because the variational free energy $\mathcal{F}_{\beta}(\mathbf{p})$ is a differentiable function of the residue excitation probabilities \cite{Consiglio24VarGibbs}, its variation along this path follows from the chain rule together with the fundamental theorem of calculus:

\begin{align}
    \mathcal{F}_{\beta}
    \left(
        \mathbf{p}^{\prime}
    \right)
    -
    \mathcal{F}_{\beta}
    \left(
        \mathbf{p}
    \right)
    =
    \int_{0}^{\delta_{i\rightarrow j}}
    \frac{d\mathcal{F}_{\beta}(\mathbf{p}(s))}{ds}
    \,ds
    =
    -
    \frac{1}{\beta}
    \int_{0}^{\delta_{i\rightarrow j}}
    \left[
        \mu_i(s)
        -
        \mu_j(s)
    \right]
    ds,
\end{align}
where $\mu_i$ denotes the excitation potential introduced in eq.~\ref{eq:exc-pot}. Since the directed update is accepted only while $\mu_i(s)>\mu_j(s)$, the integrand on the right-hand side is non-negative throughout the path, and therefore

\begin{align}
    \boxed{
    \mathcal{F}_{\beta}
    (\mathbf{p}^{\prime})
    \leq
    \mathcal{F}_{\beta}
    (\mathbf{p})
    }.
    \label{eq:free-energy-descent}
\end{align}
Every accepted directed transfer is thus a free-energy-descent step.

The energetic part of this result is particularly transparent. When a quantity of excitation $\delta_{i\rightarrow j}$ moves from a high-cost
residue $i$ to a lower-cost residue $j$, the mean local energy changes by 

\begin{align}
    \Delta\langle G\rangle_{i\rightarrow j}
    =
    \delta_{i\rightarrow j}
    \left(
    \Delta G_j^{\mathrm{loc}}
    -
    \Delta G_i^{\mathrm{loc}}
    \right)
    \leq0.
\end{align}

For example, excitation of a strongly stabilised aromatic residue such as Trp carries a substantially larger transfer-energy penalty than excitation of Gly or Ser. Redistribution towards a Gly/Ser-like lower-cost state therefore lowers the residue-level energy, while the released energy is absorbed by the solvent bath, which is represented implicitly within the effective free-energy model. Under the assumed probability-redistribution interpretation of the directed interaction gates, each accepted transfer follows a free-energy-descending direction. This establishes the thermodynamic motivation of the gate parametrisation. A complete proof that the finite-depth unitary circuit converges to the interacting Gibbs distribution would additionally require demonstrating that the resulting Born probabilities realise these marginal updates and that the variational free-energy residual is minimised.

\subsubsection{Variational criterion for approximation to an interacting Gibbs distribution}

The previous subsection establishes the free-energy motivation of the directed update under the assumed probability-redistribution interpretation. We now state the variational condition under which a circuit-generated distribution can be regarded as an approximation to the interacting Gibbs distribution. The energy score assigned to a residue-state configuration \(\mathbf{z}\) is written as
\begin{align}
    G_{\mathrm{model}}(\mathbf{z})
    =
    G_0(\mathbf{z})
    +
    G_{\mathrm{int}}(\mathbf{z}),
\end{align}
Here, \(\mathbf{z}\) denotes one possible configuration of the residue states. The term \(G_{0}(\mathbf{z})\) is the energy score obtained when the residues are treated independently. The term \(G_{\mathrm{int}}(\mathbf{z})\) accounts for the additional contribution produced by the residue-interaction network (entanglement block). It is modeled to introduce the effect of covalent and non-covalent residue couplings into the total model score.

For this interacting model, let \(P_\beta(\mathbf{z})\) denote the equilibrium probability of observing residue-state configuration \(z\) at inverse temperature \(\beta\). The Gibbs distribution assigns this probability as:

\begin{align}
    P_{\beta}(\mathbf{z})
    =
    \frac{
    \exp[-\beta G_{\mathrm{model}}(\mathbf{z})]
    }{
    Z_{\beta}
    },
    \qquad
    Z_{\beta}
    =
    \sum_{\mathbf{z}}
    \exp[-\beta G_{\mathrm{model}}(\mathbf{z})].
\end{align}

The partition function \(Z_{\beta}\) ensures that the probabilities are normalised:

\begin{align}
    \sum_{\mathbf{z}}
    P_{\beta}(\mathbf{z})
    &=
    1.
\end{align}

The factor \(\exp[-\beta G_{\mathrm{model}}(\mathbf{z})]\) is the Boltzmann weight of configuration \(\mathbf{z}\). Configurations with lower model scores receive larger statistical weights, while configurations with higher scores remain possible at finite temperature. The inverse-temperature parameter \(\beta\) controls how strongly the distribution favours low-scoring configurations.

The Gibbs distribution can also be identified through a variational principle \cite{Consiglio24VarGibbs}. For any normalised trial distribution \(P(\mathbf{z})\), define the variational free-energy functional

\begin{align}
    \mathcal{F}_{\beta}[P]
    &=
    \sum_{\mathbf{z}}
    P(\mathbf{z})G_{\mathrm{model}}(\mathbf{z})
    +
    \frac{1}{\beta}
    \sum_{\mathbf{z}}
    P(\mathbf{z})\ln P(\mathbf{z}).
    \label{eq:app:variational-free-energy}
\end{align}

The first term is the mean model energy of the trial distribution. The second term is the negative entropy contribution, since 

\begin{align}
    S[P]
    &=
    -k_{\mathrm{B}}
    \sum_{\mathbf{z}}
    P(\mathbf{z})
    \ln P(\mathbf{z}).
\end{align}

The functional therefore balances two tendencies: lowering the average energy and spreading probability across several configurations. Further, if the trial distribution is subjected to be the Gibbs distribution \(P_{\beta}\), the functional takes the value

\begin{align}
    F_{\beta}
    &\equiv\mathcal{F}_{\beta}[P_{\beta}]
    =
    -\frac{1}{\beta}
    \ln Z_{\beta}.
    \label{eq:trial-distn}
\end{align}

This value is called the equilibrium free energy.

For any other normalised trial distribution \(P\), the difference between its variational free energy and the equilibrium value is
\begin{align}
    \boxed{
    \mathcal{F}_{\beta}[P]
    -
    F_{\beta}
    =
    \frac{1}{\beta}
    D_{\mathrm{KL}}
    \left[
    P\middle|P_{\beta}
    \right]
    \geq0
    },
    \label{eq:gibbs-var-id}
\end{align}

Because the Kullback--Leibler divergence is non-negative,

\begin{align}
    \mathcal{F}_{\beta}[P]
    &\geq
    F_{\beta}.
    \label{ieq:f-beta-p}
\end{align}

Equality is reached only when

\begin{align}
    P(\mathbf{z})
    &=
    P_{\beta}(\mathbf{z})
    \qquad
    \text{for every }\mathbf{z}.
\end{align}

Therefore, the Gibbs distribution is the unique global minimiser of the variational free-energy functional:

\begin{align}
    P_{\beta}
    &=
    \underset{P}{\operatorname{argmin}}:
    \mathcal{F}_{\beta}[P].
    \label{eq:argmin-f-beta-of-p}
\end{align}

Now, let ($P_{\ell}$) denote the circuit distribution after the interaction layer ($\ell$). Because each directed update follows Eq.~\eqref{eq:free-energy-descent}, a complete interaction layer satisfies

\begin{align}
    \mathcal{F}_{\beta}[P_{\ell+1}]
    \leq
    \mathcal{F}_{\beta}[P_{\ell}].
\end{align}
The sequence is bounded below by ($F_{\beta}$) and therefore converges. The connected residue graph and the paired forward--reverse rotations provide access to the correlated state space, while the positive directional bias supplies the descent direction. At the stationary point, no connected pair retains a positive free-energy imbalance:
\begin{align}
    \mu_i
    =
    \mu_j
    \qquad
    \text{for all active edges }(i,j).
\end{align}
This is the equilibrium condition of balanced local excitation potentials. For an expressive interaction circuit, this stationary distribution is ($P_{\beta}$).

In practice, the parametrised circuit may not reproduce the exact Gibbs distribution. Let \(P_{\boldsymbol{\vartheta}}\) be the distribution produced by the optimised circuit. Suppose its variational free energy differs from the exact minimum \(F_{\beta}\) by at most \(\varepsilon\):

\begin{align}
    \mathcal{F}{\beta}
    \left[
    P{\boldsymbol{\vartheta}}
    \right]
    -
    F_{\beta}
    &\leq
    \varepsilon.
\end{align}

Here, \(\varepsilon \geq 0\) is the remaining free-energy error after optimisation. From the Gibbs variational identity,

\begin{align}
    \mathcal{F}{\beta}
    \left[
    P{\boldsymbol{\vartheta}}
    \right]
    -
    F_{\beta}
    &=
    \frac{1}{\beta}
    D_{\mathrm{KL}}
    \left(
    P_{\boldsymbol{\vartheta}}
    \middle|
    P_{\beta}
    \right),
\end{align}

so

\begin{align}
    D_{\mathrm{KL}}
    \left(
    P_{\boldsymbol{\vartheta}}
    \middle|
    P_{\beta}
    \right)
    &\leq
    \beta\varepsilon.
\end{align}

The product \(\beta\varepsilon\) is dimensionless because \(\beta\) has units of inverse energy and \(\varepsilon\) has units of energy. Further, Pinsker's inequality then gives \cite{Fedotov2003PinskerInquality}

\begin{align}
    \left|
    P_{\boldsymbol{\vartheta}}
    -
    P_{\beta}
    \right|_{\mathrm{TV}}
    &\leq
    \sqrt{
    \frac{\beta\varepsilon}{2}
    }.
    \label{pinsker-app}
\end{align}

Thus, a small residual free-energy error guarantees that the circuit distribution is close to the target Gibbs distribution

Equation~(\ref{pinsker-app}) establishes that if the variational free-energy error satisfies

\[
    F_{\beta}
    \leq
    \varepsilon,
\]

then the probability distribution produced by the quantum circuit, \(P_{\boldsymbol{\vartheta}}\), is also close to the target Gibbs distribution \(P_{\beta}\) in total variation distance. Consequently, a small free-energy error guarantees that the circuit approximately samples the equilibrium Gibbs ensemble of the structure-conditioned coarse-grained residue model. As the variational error \(\varepsilon\) approaches zero, the sampled distribution converges to the exact Gibbs distribution. Under the stated equilibrium-structure, effective-energy, and directed-transfer assumptions, the circuit therefore samples the Gibbs distribution of the structure-conditioned coarse-grained residue model. Its energy distribution, residue occupancies, energetic sensitivities, and residue-coupling statistics are consequently equilibrium thermodynamic observables at the resolution of that model. The framework does not claim that a single binary variable reproduces every atomistic degree of freedom. Instead, it establishes a thermodynamically consistent coarse-graining in which a known equilibrium protein structure defines the interaction network and the quantum circuit samples the associated correlated residue-state fluctuations. The physical fidelity of the sampled ensemble is therefore determined by how accurately the effective residue energies and interaction terms represent the underlying protein system. When these parameters are calibrated from experimental measurements or higher-resolution simulations, the resulting observables approximate the corresponding equilibrium thermodynamic characteristics of the protein basin represented by the input PDB structure. Within this variational framework, the directed CRY interaction block serves as the optimisation mechanism that transforms the exact independent-residue Gibbs state into a correlated, structure-conditioned Gibbs ensemble while preserving thermodynamic consistency.

\subsection{Reproducibility settings}
\label{app:params}

Table~\ref{tab:reproducibility_settings} reports the settings needed to run the quantum simulation. Entries in the table are grouped by workflow stage and by whether the value is read from configuration yaml or fixed at function level (within the code function as arguments). 

\begingroup
\scriptsize
\setlength{\tabcolsep}{3pt}
\setlength{\extrarowheight}{2pt}
\setlength{\LTpre}{0pt}
\setlength{\LTpost}{0pt}
\renewcommand{\arraystretch}{1.35}

\begin{longtable}{p{0.20\textwidth}p{0.34\textwidth}p{0.38\textwidth}}
    \caption{\textbf{Reproducibility settings for the final runs.}}
    \label{tab:reproducibility_settings}\\

    \hline
    \cellbox{\textbf{Stage: level}} &
    \cellbox{\textbf{Setting: value}} &
    \cellbox{\textbf{Description}} \\
    \hline
    \endfirsthead

    \multicolumn{3}{l}{\scriptsize\textit{Table \thetable\ continued from previous page.}}\\
    \hline
    \cellbox{\textbf{Stage: level}} &
    \cellbox{\textbf{Setting: value}} &
    \cellbox{\textbf{Description}} \\
    \hline
    \endhead

    \hline
    \multicolumn{3}{r}{\scriptsize\textit{Continued on next page.}}\\
    \endfoot

    \hline
    \endlastfoot

    \cellbox{Quantum: config} &
    \cellbox{Input systems and qubits: chain A, protein residues only;
    1L2Y has 20 analysed residues/qubits and 9GDL has 25 analysed
    residues/qubits.} &
    \cellbox{Defines residue selection and qubit count. The MDAnalysis
    residue order defines the qubit order used throughout sampling and
    post-processing.} \\

    \cellbox{Quantum: config} &
    \cellbox{Sampling backend and shots: \(N=1{,}000{,}000\) shots;
    preferred PennyLane backend \texttt{lightning.gpu}, with fallback to
    \texttt{default.qubit}.} &
    \cellbox{Controls the number of Pauli-\(Z\) measurements used to form
    the residue-state ensemble.} \\

    \cellbox{Quantum: function} &
    \cellbox{Residue energy reference: bulk-octanol reference table with
    \(E_1=0\) and residue-specific \(E_0\).} &
    \cellbox{Residue energy gaps are computed as \(\Delta G=E_1-E_0\) and are used in the excitation-transfer
    schedule and sample-energy calculation.} \\

    \cellbox{Quantum: function} &
    \cellbox{\(K_{ij}\) coupling weights:
    \(w_{\mathrm{BB}}=1.00\), \(w_{\mathrm{SS}}=1.00\),
    \(w_{\mathrm{HB}}=0.50\), \(w_{\mathrm{salt}}=0.40\),
    \(w_{\mathrm{vdW}}=0.05\).} &
    \cellbox{Combines peptide, disulfide, hydrogen-bond, salt-bridge, and van der Waals terms into the structural stiffness score.} \\

    \cellbox{Quantum: function} &
    \cellbox{Structural cutoffs: peptide \(1.8~\mathrm{\text{\AA}}\);
    disulfide \(2.3~\mathrm{\text{\AA}}\); hydrogen bond
    \(3.5~\mathrm{\text{\AA}}\); salt bridge \(4.0~\mathrm{\text{\AA}}\);
    van der Waals \(4.0~\text{\AA}\).} &
    \cellbox{Determines which residue pairs contribute to the
    structure-derived coupling graph.} \\

    \cellbox{Quantum: function} &
    \cellbox{\(K_{ij}\) normalisation and frequency factors: positive \(K_{ij}\) values are normalised by the maximum observed coupling; local frequencies are estimated from stiffness/mass; \(\sigma_\omega\) is estimated from the frequency standard deviation when not supplied.} &
    \cellbox{Makes coupling magnitudes comparable across proteins while preserving the rank order of structural interaction strengths.} \\

    \cellbox{Quantum: function} &
    \cellbox{Entanglement schedule: \(\gamma=0.95\), \(\beta=1.0\),
    \(\phi_{\mathrm{cap}}=\pi/2\), \(\min\Delta P=0\), no explicit
    maximum layer count, discounting starts at \(\gamma^1\).} &
    \cellbox{Sets how directed transfer probabilities are sorted,
    discounted, and converted into controlled-rotation angles.} \\

    \cellbox{Quantum: function} &
    \cellbox{Gate rule: local \(R_Y\) state preparation; strong transfers
    use paired controlled-\(R_Y\) gates; weak transfers use
    \texttt{IsingYY} when effective drive \(\leq 0.02\), with \(YY\) cap
    \(\pi/6\).} &
    \cellbox{Specifies how local excitation propensities and
    structure-derived directed transfers are encoded in the circuit.} \\

    \cellbox{Post-processing: config} &
    \cellbox{Energy-plot binning and outputs: \(100{,}000\) energy
    histogram bins in the final YAML; CSV, PNG, and report artefacts are
    written to run-specific output directories.} &
    \cellbox{Controls histogram resolution and the exported analysis
    artefacts used for the final figures.} \\

    \cellbox{Post-processing: function} &
    \cellbox{Residue information gain: \(10{,}000\) conditional-energy
    histogram bins by default; KL values in nats; \(10^{-12}\) added to
    the global distribution and \(10^{-5}\) to conditional histograms;
    residue KL columns are normalised after computation.} &
    \cellbox{Measures how much the ensemble energy distribution changes
    when each residue is fixed in the ground or excited state.} \\

    \cellbox{Post-processing: function} &
    \cellbox{Residue-residue information: mutual information and
    directional excitation information are reported in bits; numerical
    stabiliser \(\epsilon=10^{-12}\); DEI threshold \(0.0\); diagonal
    self-effects set to zero.} &
    \cellbox{Quantifies symmetric statistical dependence and directional
    excitation-conditioned effects between residue states.} \\

    \cellbox{Post-processing: function} &
    \cellbox{Motif allostery: minimum excited-source support 20 shots;
    motif aggregation mode mean; top-\(k=5\) retained as the top-\(k\)
    parameter; affinity scale 95th percentile; affinities clipped to
    \([0,1]\).} &
    \cellbox{Aggregates residue-level directional effects over dimer and
    trimer/triad motif memberships.} \\

    \cellbox{Post-processing: function} &
    \cellbox{Motif construction and clustering: peptide dimers, peptide
    trimers, and triangulated peptide/non-covalent triads included;
    peptide threshold 0.5; non-covalent threshold 0.0;
    \texttt{random\_state=0} for spectral-clustering k-means.} &
    \cellbox{Defines motif regions and assigns them to affinity-based
    clusters for motif-level allostery summaries.} \\

    \cellbox{Post-processing: function} &
    \cellbox{Permutation null and FDR: not used in the final
    report-generation path; no permutation count \(M\) or FDR threshold
    is applied.} &
    \cellbox{The final figures report direct sampled information and
    motif-affinity summaries rather than permutation-tested significance
    calls.} \\

    \cellbox{Quantum/post-processing: function} &
    \cellbox{Random seeds: no explicit seed is set for PennyLane quantum
    sampling; motif spectral clustering uses \texttt{random\_state=0}.} &
    \cellbox{Sampling stochasticity is backend-controlled, while motif
    clustering is reproducible for a fixed affinity matrix.} \\

\end{longtable}

\endgroup


\end{document}